\DocumentMetadata{}
\documentclass[sigconf, 10pt]{acmart}
\usepackage{tikz}

\usepackage{filecontents}

\usepackage{setspace, amsmath, url, lscape, algorithmic, multirow, pslatex, listings, verbatim, alltt, amsfonts, wrapfig, boxedminipage, color, bookmark}
\usepackage[vlined,linesnumbered,ruled,boxed]{algorithm2e}

\usepackage{subcaption}

\DeclareFontFamily{\encodingdefault}{\ttdefault}{\hyphenchar\font=`\-}
\newcommand\ProcessThreeDashes{\llap{\color{cyan}\mdseries-{-}-}}

\usepackage{paralist}
\usepackage{epsfig}
\let\labelindent\relax
\usepackage[inline]{enumitem}

\usepackage{xcolor}

\usepackage{listings}

\newcommand\YAMLcolonstyle{\color{red}\mdseries}
\newcommand\YAMLkeystyle{\color{black}\bfseries}
\newcommand\YAMLvaluestyle{\color{blue}\mdseries}

\makeatletter

\newcommand\language@yaml{yaml}

\expandafter\expandafter\expandafter\lstdefinelanguage
\expandafter{\language@yaml}
{
  keywords={true,false,null,y,n},
  keywordstyle=\color{darkgray}\bfseries,
  basicstyle=\ttfamily,                                 
  sensitive=false,
  comment=[l]{\#},
  morecomment=[s]{/*}{*/},
  commentstyle=\color{purple}\ttfamily,
  stringstyle=\YAMLvaluestyle\ttfamily,
  moredelim=[l][\color{orange}]{\&},
  moredelim=[l][\color{magenta}]{*},
  moredelim=**[il][\YAMLcolonstyle{:}\YAMLvaluestyle]{:},   
  morestring=[b]',
  morestring=[b]",
  literate =    {---}{{\ProcessThreeDashes}}3
                {>}{{\textcolor{red}\textgreater}}1     
                {|}{{\textcolor{red}\textbar}}1 
                {\ -\ }{{\mdseries\ -\ }}3,
}

\lst@AddToHook{EveryLine}{\ifx\lst@language\language@yaml\YAMLkeystyle\fi}
\makeatother

\DeclareMathDelimiter{(}{\mathopen} {operators}{"28}{largesymbols}{"00}
\DeclareMathDelimiter{)}{\mathclose}{operators}{"29}{largesymbols}{"01}

\newcommand{\qedd}{\nobreak \ifvmode \relax \else
      \ifdim\lastskip<1.5em \hskip-\lastskip
     \hskip1.5em plus0em minus0.5em \fi \nobreak
      \vrule height0.75em width0.5em depth0.25em\fi}

\newcommand{\eg}{{e.g., }}

\newcommand{\comments}[1]{}
\newcommand\hl{\bgroup\markoverwith
  {\textcolor{yellow}{\rule[-.5ex]{2pt}{2.5ex}}}\ULon}

\copyrightyear{2024}
\acmYear{2024}
\setcopyright{acmlicensed}\acmConference[SoCC '24]{ACM Symposium on Cloud Computing}{November 20--22, 2024}{Redmond, WA, USA}
\acmBooktitle{ACM Symposium on Cloud Computing (SoCC '24), November 20--22, 2024, Redmond, WA, USA}
\acmDOI{10.1145/3698038.3698552}
\acmISBN{979-8-4007-1286-9/24/11}

\begin{document}

\title{Streamlining Cloud-Native Application Development and Deployment with Robust Encapsulation}

\author{ Pawissanutt Lertpongrujikorn}
\email{pawissanutt.lertpongrujikorn@unt.edu}
\orcid{0009-0003-4106-2347}
\authornote{These authors contributed equally to this work}
\affiliation{
  \institution{HPCC Lab., Uni. North Texas}
  \city{Denton}
  \state{Texas}
  \country{USA}
}

\author{Hai Duc Nguyen}
\email{ndhai@cs.uchicago.edu}
\orcid{0000-0003-4177-0493}
\affiliation{
  \institution{Argonne National Laboratory and University of Chicago}
  \city{Chicago}
  \state{Illinois}
  \country{USA}
}
\authornotemark[1]

\author{Mohsen Amini Salehi}
\email{mohsen.aminisalehi@unt.edu}
\orcid{0000-0002-7020-3810}
\affiliation{
  \institution{HPCC Lab, Uni. North Texas}
  \city{Denton}
  \state{Texas}
  \country{USA}
}


\begin{abstract}


Current Serverless abstractions (\eg FaaS) poorly support non-functional requirements (e.g., QoS and constraints), are provider-dependent, and are incompatible with other cloud abstractions (e.g., databases). As a result, application developers have to undergo numerous rounds of development and manual deployment refinements to finally achieve their desired quality and efficiency.
In this paper, we present Object-as-a-Service (OaaS)---a novel serverless paradigm that borrows the object-oriented programming concepts to encapsulate business logic, data, and non-functional requirements
into a single deployment package, thereby streamlining provider-agnostic cloud-native application development. We also propose a declarative interface for the non-functional requirements of applications that relieves developers from daunting refinements to meet their desired QoS and deployment constraint targets. We realized the OaaS paradigm through a platform called Oparaca and evaluated it against various real-world applications and scenarios.
The evaluation results demonstrate that Oparaca can enhance application performance by 60$\times$ and improve reliability by 50$\times$ through latency, throughput, and availability enforcement---all with remarkably less development and deployment time and effort.

\end{abstract}


\maketitle


\section{Introduction}
\label{sec:introduction}

Function-as-a-Service (FaaS), or serverless computing, has emerged as a transformative paradigm in cloud computing, redefining how businesses and individuals develop and deploy applications. Unlike traditional virtualized infrastructure (e.g., virtual machines), FaaS enables on-demand code execution in response to events, eliminating the need to manage servers or underlying infrastructure. Developers leverage FaaS through high-level abstractions provided by cloud platforms, allowing them to focus on writing and running functions rather than managing complex systems, thereby significantly enhancing productivity.

\begin{figure}
    \centering
    \begin{subfigure}{0.5\textwidth}
        \centering
        \includegraphics[scale=0.6]{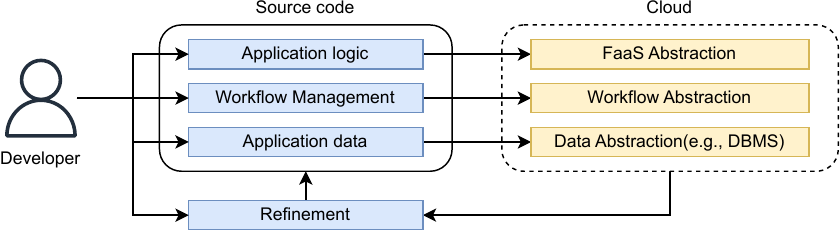}
        \caption{FaaS-based application}
        \label{fig:current-cloud-apps}
    \end{subfigure}
    \begin{subfigure}{0.5\textwidth}
        \centering
        \includegraphics[scale=0.6]{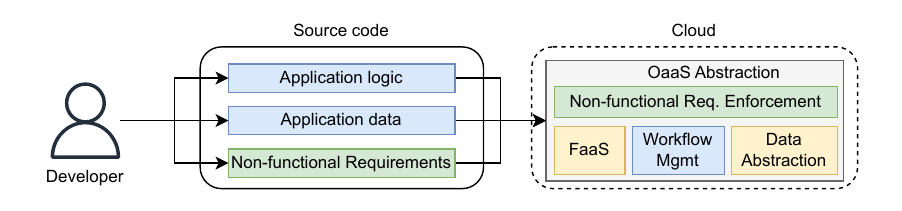}
        \caption{OaaS-based application}
        \label{fig:oaas-apps}
    \end{subfigure}
    \caption{
    OaaS extends the FaaS abstraction to encapsulate everything into a single deployment with built-in non-functional enforcement, boosting productivity and efficiency.
    }
    \label{fig:intro}
\end{figure}

  

Unfortunately, beyond hiding complexity, FaaS plays a very limited role in other aspects. Primarily, FaaS functions only offer resource and computation abstraction, which is insufficient for a complete application deployment \cite{jonas2019cloud, shafiei2022serverless, denninnart2023efficiency}. Developers must rely on additional cloud services, such as databases \cite{aws-s3, amazon-rds} and orchestrators \cite{aws-sf, azure-df}, to manage application states and workflows.
Yet, there are limited integration supports among these abstractions. In stateful applications, for example, FaaS performance depends on data locality, but current FaaS implementations provide no means to help FaaS functions cooperate with cloud data abstractions in this regard, negatively affecting productivity and efficiency. Furthermore, the FaaS abstraction is implemented by cloud providers, who usually prioritize system metrics, such as resource utilization, which can result in unpredictable and uncontrollable quality degradation on the application side \cite{nguyen2023storm}. This leads to counterproductive interactions between developers and the cloud, such as over-provisioning and over-subscription \cite{pandey2023optimizing, zhou2022aquatope}. These limitations force developers to manually reconfigure their deployments through multiple rounds of refinement for their \textit{non-functional requirements} (e.g., Quality of Service, a.k.a. QoS). The process lacks proper guidance and relies heavily on resource-domain expertise and experience \cite{kuhlenkamp2019evaluation}, making cloud application development and deployment complex and costly \cite{nguyen2019real} (see Figure \ref{fig:current-cloud-apps}).

\begin{figure*}[htb!]
    \centering
    \begin{subfigure}{0.6\textwidth}
        \centering
        \includegraphics[height=1.9in, scale=0.6, trim={10, 50, 10, 50}, clip]{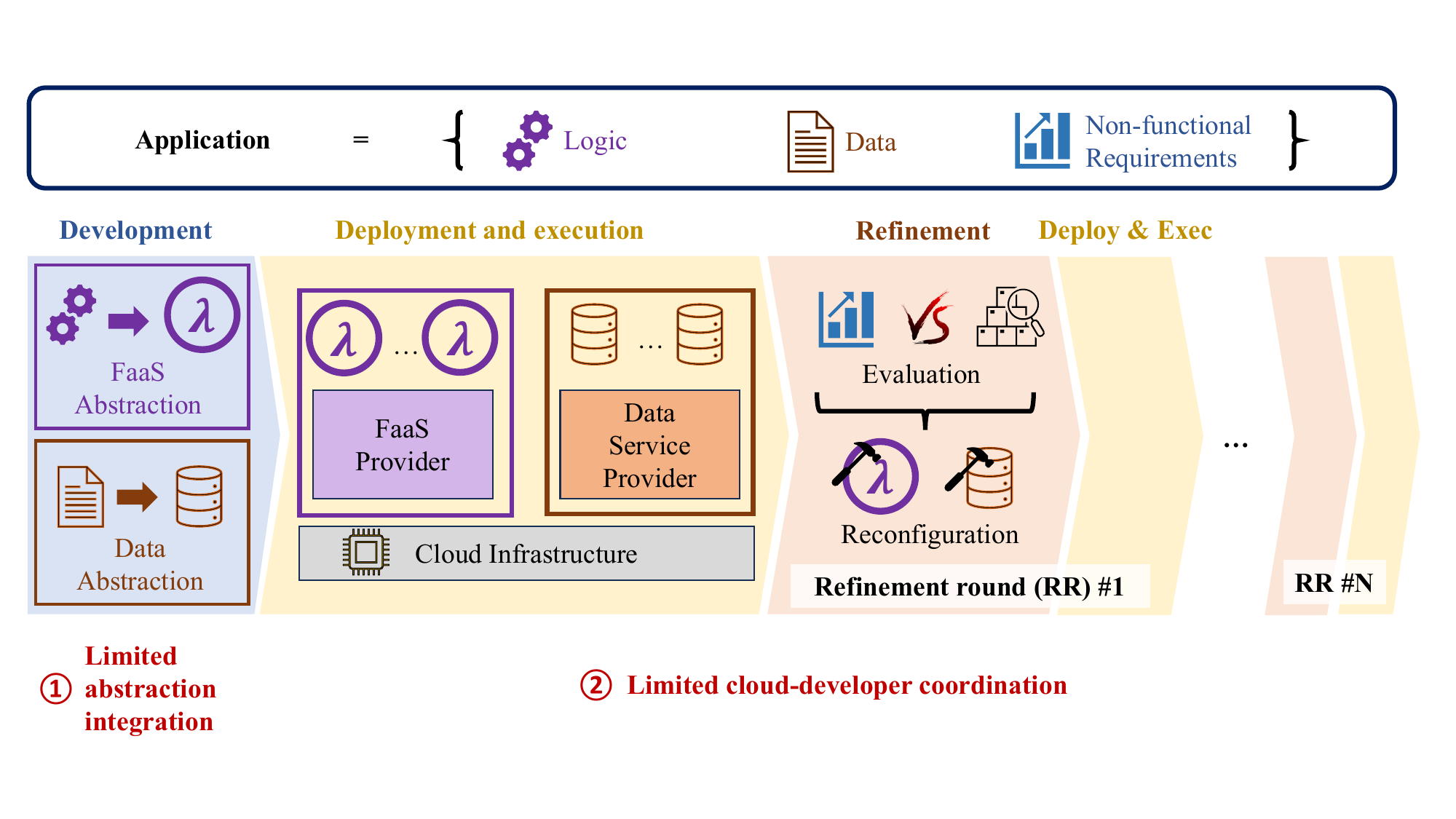}
        \caption{Current FaaS-based application life cycle and its problems}
        \label{fig:cloud-problems}
    \end{subfigure}%
    \begin{subfigure}{0.4\textwidth}
        \centering
        \includegraphics[height=1.9in, scale=0.6, trim={10 50 10 50},clip]{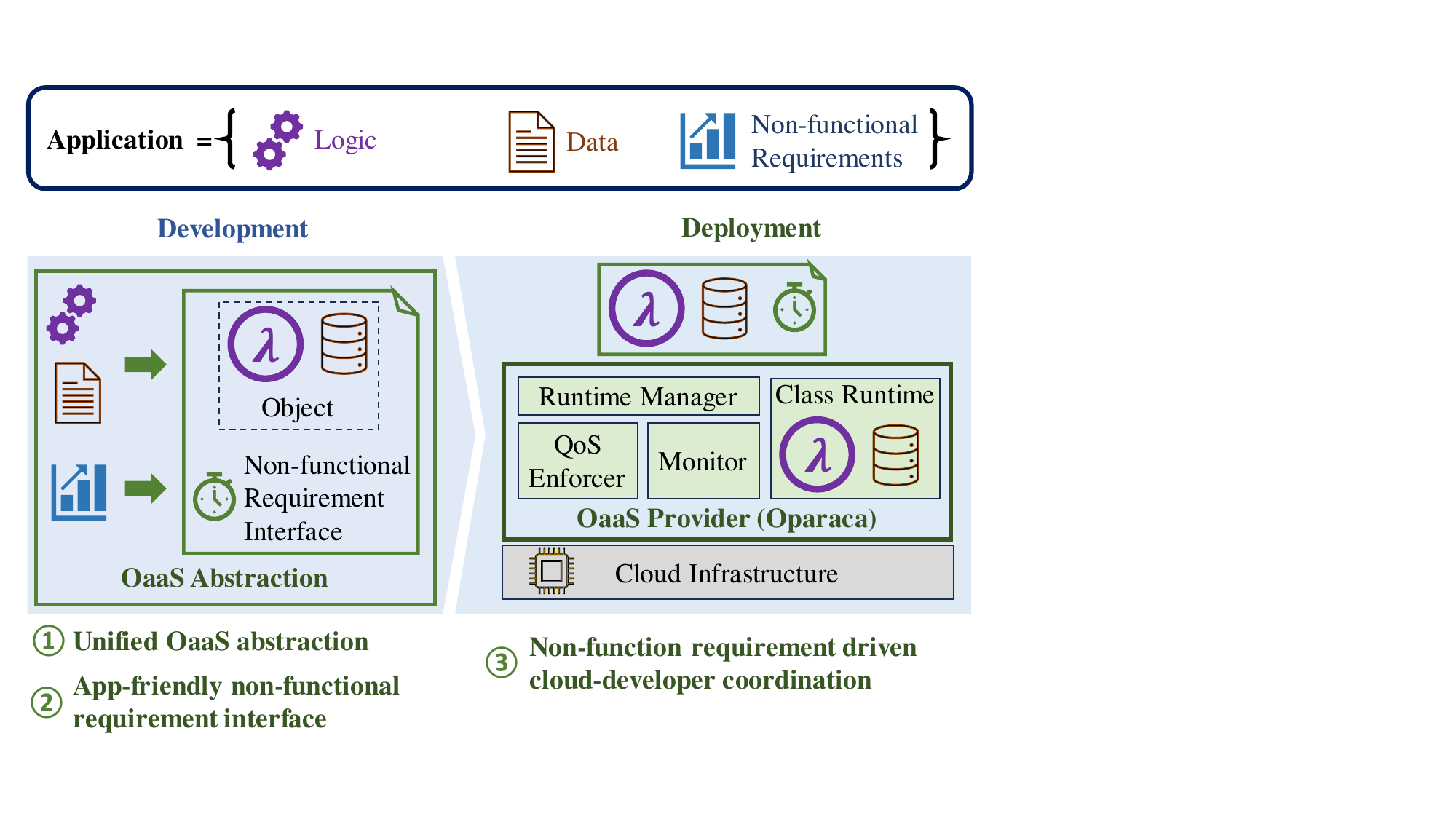}
        \caption{OaaS and non-functional requirements abstraction}
        \label{fig:oaas-ideas}
    \end{subfigure}
    \caption{Limited non-functional support from the cloud introduces repeated and complex refinement processes for quality of service. In contrast, the OaaS abstraction outsources the refinement to the cloud provider with a well-defined set of information and refinement objectives. Thus making the application development and deployment more productive.}
    \label{fig:problem-approach}
\end{figure*}


To resolve the problem, we propose Object-as-a-Service (OaaS) abstraction, a new cloud computing paradigm that borrows the concepts of object-oriented programming to let the cloud applications include their logic (i.e., functions), data (i.e., state), and non-functional requirements into a single deployment package (see Figure \ref{fig:oaas-apps}). The OaaS abstraction allows developers to unify the application functionality implementations into single packages, eliminating the multi-abstraction barriers that hinder productivity and efficiency. The abstraction also comes with a ``non-functional requirement interface'' that allows applications to declare the expected QoS requirements and constraints as high-level and measurable metrics. Developers can use the interface to express their requirements, and then the cloud provider will enforce them automatically, removing the need for repeated refinements.
This unlocks new opportunities for cloud optimization.
With OaaS deployment, the cloud providers are given a clear set of optimization objectives (from the ``non-functional interface'') with a rich set of information (from the ``object'') so that they will know the right direction to optimize their system metrics while still be able to meet their customer requirements.



We implement Oparaca, an open-source platform that realizes ideas of OaaS to simplify application deployment. Oparaca integrates various object deployment and management approaches, each specialized for specific object structures and requirement combinations, and then optimizes them for different deployment scenarios. Thus, making object deployment portable and efficient.
We systematically evaluate Oparaca versus state-of-the-art solutions through various experiments on real Cloud testbeds. We found that 
Oparaca often outperforms the other baseline approaches in efficiently meeting the targeted performance objectives (e.g., throughput) of multiple services for different application types. In contrast, other baselines struggle due to resource contention issues arising from the lack of service-specific awareness of the targeted performance objective.

In sum, the contributions of this research are as follows:
\begin{itemize} [leftmargin=*, itemsep=0pt, topsep=2pt]
    \item Object-as-a-Service (OaaS) abstraction that exploits the Object-oriented programming concepts to encapsulate functional and non-functional requirements into one deployment package, enhancing application development and deployment productivity.
    \item A non-functional requirements interface that lets developers express their non-functional requirements in a human-friendly and measurable manner, thus enabling application portability and opportunities for cooperative cloud-application interactions.
    \item Oparaca -- an OaaS prototype implementation that enables simple, scalable, and QoS-aware applications development.
    \item Evaluating and analyzing the Oparaca from the QoS enforcement, efficiency, and productivity perspectives. By leveraging the OaaS abstraction, applications can improve their performance by 60$\times$ and availability by 50$\times$ with much less deployment time and effort. 
\end{itemize}

The remainder of the paper is organized as follows. In Section~\ref{sec:problem}, we present the issues of the current FaaS abstraction that make delivering efficient solutions complicated and expensive. We show how OaaS abstraction resolves these issues in Section \ref{sec:approach} and with more details in Sections \ref{sec:architecture}.
Section \ref{sec:evaluation} evaluates Oparaca against state-of-the-art solutions under various scenarios. We briefly present related work in Section \ref{sec:related-work} and summarize the paper in Section \ref{sec:conclusion}. 

\section{Motivation and Problem Statement}
\label{sec:problem}

Figure \ref{fig:cloud-problems} shows a typical life cycle of a FaaS-based application that consists of three primary phases:
\begin{itemize}[leftmargin=*, itemsep=0pt, topsep=2pt]
    \item \textit{Development}: The application developers design suitable logic/algorithms and data structure for the application and then encapsulate them into \textit{separated deployment packages}
    (e.g., FaaS deployments, database schema, etc.). 
    \item \textit{Deployment and Execution}: cloud provider receives deployment packages and executes them separately across their infrastructure through \textit{service providers} (e.g., FaaS and data service providers). Each one is implemented and optimized specifically for a cloud service abstraction (e.g., AWS Lambda for FaaS abstraction, MySQL for Relational DB, etc.).
    \item \textit{Refinement}: Applications typically have non-functional requirements specifying their QoS (e.g., desired throughput, availability, etc.) and execution constraints (e.g., budget, Carbon footprint, jurisdiction, etc.). To ensure these requirements, developers have to evaluate them against monitored data. Refinement, which includes reconfiguring and redeploying the application packages, is needed if any of these requirements fail to be met.
\end{itemize}

In practice, the refinement phase consists of multiple rounds of reconfiguration and deployments that cost a lot of time and effort \cite{kuhlenkamp2019evaluation, nguyen2019real}. This is because \textit{current FaaS implementations and their supportive services offer limited supports that make it complicated and expensive to deliver efficient cloud applications}. The problem manifests in many aspects of cloud application life cycles, as outlined below. 

\begin{figure}
    \centering
    \subfloat[Locality Impact]{\includegraphics[width=0.23\textwidth]{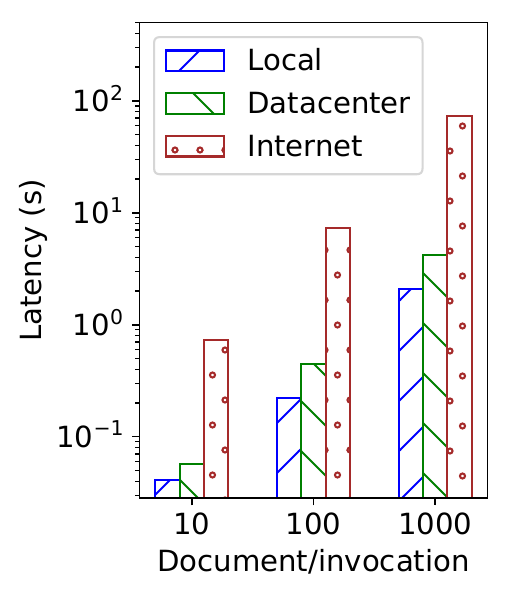}\label{fig:locality-faas}} 
    \hfill
    \subfloat[Configuration Impact]{\includegraphics[width=0.23\textwidth]{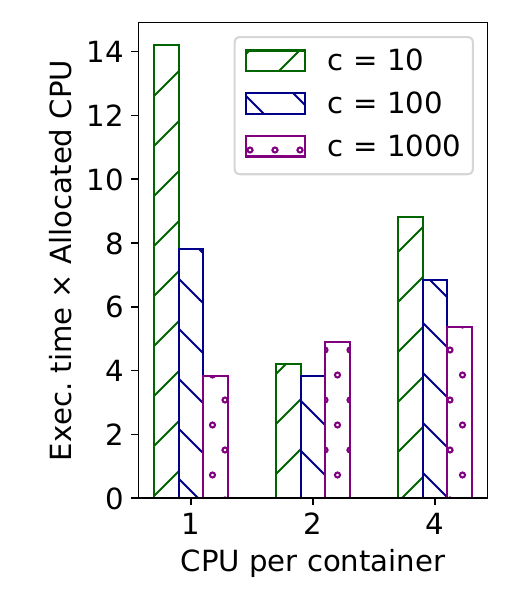}\label{fig:config-faas}}
    \caption{FaaS limitations (a) prevent applications from exploiting locality for performance and (b) complicate deployment refinement.}
    \label{fig:faas-limits}
\end{figure}



\vspace{2mm}
\noindent\textbf{Limited Abstraction Integration.} 
FaaS applications are formed based on FaaS functions and additional cloud services, such as databases and workflow orchestrators. However, these abstractions typically operate independently. This independence creates challenges, as even a single functionality may involve multiple abstractions but lacks the capability to integrate them effectively. In stateful applications, for example, FaaS functions typically need access to an external database to read and update its state. This makes the application performance strongly depend on efficient data transmission between the FaaS invocations and the database.
Figure \ref{fig:locality-faas} illustrates the end-to-end latency of a FaaS invocation chain modifying JSON documents from a database (see Section \ref{sec:evaluation-setup}), varying the database's location relative to the function's containers and the number of JSON documents modified. Clearly, executing a function on the same machine as the database (``Local'') is significantly faster than running the function from a different machine within the same data center ($2\times$) or across the Internet ($35\times$). The latency difference increases substantially as more data is transmitted (e.g., the ``Datacenter'' to ``Local'' latency ratio increases by $1.4\times$ when we increase the number of modified documents per invocation from 10 to 1000). Based on the results, FaaS should leverage data locality by dispatching functions close to the database to minimize end-to-end latency. Unfortunately, current FaaS abstractions lack support for such cross-abstraction optimization, forcing developers to create their own solutions, which is both challenging and effort-consuming \cite{carreira2019cirrus, jia2021nightcore, mahgoub2021sonic, muller2020lambada, pu2019shuffling, yu2023following}.

\vspace{2mm}
\noindent\textbf{Limited Cloud-Developer Coordination.}
Developers refine application deployments through primitive resource-domain settings, like per-container CPU allocation. On the other hand, non-functional requirements are typically measured and evaluated using application-domain metrics, such as throughput, latency, and monetary cost. Translating these requirements into effective FaaS configurations is challenging. Figure \ref{fig:config-faas} illustrates the resource cost, measured by the average \textit{execution time} $\times$ \textit{CPU} allocated to the JSON randomization deployment under different per-container concurrency and CPU allocation configurations. Considering only one factor for cost minimization is insufficient; for instance, with one CPU per container, varying concurrency ($c$) can change costs by up to $4.3\times$, but doing so has little effect with two CPUs per container. Configuring these factors together is necessary, but there are no clear insights into how to do so.
For example, increasing concurrency ($c$) generally allows more invocations per container, reducing costs. However, setting the concurrency too high can lead to resource contention, which prolongs invocation execution and increases costs. The optimal concurrency thresholds vary with different CPU allocations. With two CPUs per container, $c = 1000$ is too high, but it works well for configurations with one or four CPUs per container.
Therefore, configuring such low-level parameters to produce a reliable and robust deployment demands significant effort and expertise (e.g., \cite{bilal2023great}), often necessitating numerous rounds of refinement \cite{kehrer2021self, lin2022fine, kuhlenkamp2019evaluation}.

Worse, the implementation and configuration of FaaS and supporting cloud abstractions are influenced by the cloud providers' objectives, potentially hindering the fulfillment of application non-functional requirements. For example, many cloud resource managements employ over-subscription \cite{everman2021improving, baset2012towards, li2023golgi}, which implicitly commits more resources to users than the cloud can actually provide for better utilization. However, this practice increases the risk of interference when multiple applications peak simultaneously, leading to uncontrollable and unpredictable QoS degradation \cite{sakalkar2020data, denninnart2021harnessing, copik2023rfaas, zhou2022aquatope}. To counteract this, many applications request more resources than they need \cite{pandey2023optimizing, aws-perf-stats}, prompting providers to oversubscribe even more aggressively \cite{kumbhare2021prediction, bhasi2022cypress, bilal2023great}. This creates a harmful cycle of overestimation and mistrust, negatively affecting both applications and the cloud infrastructure \cite{ginzburg2020serverless}.

\section{Object-as-a-Service Abstraction}
\label{sec:approach}
To establish an agile and cost-efficient application delivery, the two challenges presented in Section \ref{sec:problem} must be properly addressed. In this section, we propose solutions for each challenge and then combine them to form a novel approach for FaaS-based application development and deployment. 

\begin{table*}[ht]
    \centering
    \begin{tabular}{|l|l|l|l|}
        \hline
        \textbf{Name} & \textbf{Value Type} & \textbf{Unit} & \textbf{Definition}\\ \hline \hline
        \multicolumn{4}{|l|}{\textit{QoS Requirements}}\\ \hline
        \textbf{Throughput} & \textbf{Integer} & \textbf{Rps} & \textbf{Minimum number of invocations guaranteed to be executed per second}\\ \hline
        \textbf{Availability} & \textbf{Real} & \textbf{Percent} & \textbf{The percentage of time an object/function must be available for service.}\\ \hline
        \textbf{Locality} & \textbf{\{Local, None\}} & \textbf{N/A} & \textbf{How function invocations are dispatched with respect to object state location.}\\ \hline\hline
        \multicolumn{4}{|l|}{\textit{Deployment Constraints}}\\ \hline 
        \textbf{Persistent} & \textbf{Yes/No} & \textbf{N/A} & \textbf{Should the data associated with the object persistent}\\ \hline
        \textbf{Runtime Req.} & \textbf{Dict} & \textbf{N/A} & \textbf{Specific object runtime configuration. 
        (e.g., choice of FaaS engine) } \\ \hline
        Budget & Integer & Credit & Object deployment and operation budget. All costs must not exceed this value.\\ \hline
        Consistency & Enumerate & N/A & Object consistency model: eventual, sequential, linearization, or none. \\ \hline
        Jurisdiction & Enumerate & N/A & Candidate places to deploy an object\\ \hline
        Data Encryption & Enumerate & N/A & Specify or disable the encryption algorithm for the stored data\\ \hline
    \end{tabular}
    \caption{Potential Non-functional requirements and constraints. Those with bold font are currently supported by Oparaca.}
    \label{tab:non-functional-requirements}
\end{table*}

\subsection{Unified OaaS Abstraction} 

We extend the FaaS abstraction, called Object as a Service (OaaS), that borrows the object-oriented programming (OOP) concepts to unify application logic and data
within a single abstraction. Specifically, each application is defined as a collection of cloud objects where its data (a.k.a. state) is modeled as ``attributes'' with supported data types in current cloud data abstraction, and its logic is modeled as methods realized by serverless functions.
In this manner, OaaS abstraction alone is sufficient for the entire application development phase---eliminating the need for multiple distinct abstractions and the complexities of effectively gluing them.

OaaS also offers the notions of abstract class, inheritance, and polymorphism to establish software reuse across cloud objects, thereby reducing redundancy and enhancing development productivity at the FaaS workflow level. This is in contrast to traditional FaaS, which typically limits software reuse to the function or invocation level (e.g., through shared libraries).
Beyond these, OaaS transformation unlocks new opportunities for deployment optimizations that were previously difficult or impossible.
This is because the object abstraction provides richer information for optimization and grants the cloud greater control over the deployment to exploit them. For example, 
OaaS lets application data and logic be encapsulated and managed together under the object abstraction. Thus, OaaS can easily find the data associated with each method and proactively distribute them across the cloud database instances that are close to the deployed method, thereby minimizing the data transmission overhead.

\subsection{Non-functional Requirement Interface}

Within the OaaS abstraction, we develop a non-functional requirement interface that lets the developer express their non-functional requirements in a human-friendly manner. Through the interface, developers can declare their non-functional requirements for a whole object or even for a specific part (attribute or method) of it. The requirements are defined as high-level and measurable metrics either in the form of QoS (e.g., availability and throughput) requirements or deployment constraints (e.g., budget and jurisdiction). During the deployment, the cloud provider takes these non-functional requirements as input to its internal services and adjusts their operations to meet the requirements. The benefits are three-fold: 
\begin{itemize}[leftmargin=*, itemsep=0pt, topsep=2pt]
    \item \textit{Productivity}: applications no longer need to consider low-level resource configuration for non-functional requirements. This relieves the burden of performance optimization from their deployment process, thus improving productivity.
    \item \textit{Portability}:
    as long as the cloud provider supports OaaS, the application can rely on the object abstraction to maintain its functionality, meet its QoS and constraint expectations (via the non-functional requirement interface), and comfortably deploy across scenarios with minimal changes. 
    \item \textit{Cloud-application symbiosis}: since applications use cloud resources for execution, the common sense is that the cloud should fulfill the non-functional requirement, as it has sufficient knowledge and privilege on the underlying infrastructure.
    With the non-functional requirement interface, however, the cloud does not take this responsibility alone. Here, the interface acts as a ``glue'' to make
    a symbiosis between the cloud and the application developer.
    Specifically, the requirements declared through the interface are valuable guidelines for cloud service providers to know which optimization they should follow so as not to impact the applications negatively. On the other hand, the interface is a useful means of communication that lets the developer actually configure for performance and quality, as opposed to going through multiple rounds of playing a ``trial-and-error'' game with the cloud providers to meet the desired outcomes.
\end{itemize}


\subsection{Simplified, Refinement-Free Deployment}

Based on the ideas above, as shown in Figure \ref{fig:oaas-ideas}, we propose a novel paradigm to develop and deploy cloud applications. In this paradigm, cloud applications are modeled as a set of objects, each can be developed and deployed independently. An object can possess deployment constraints and QoS requirements declared through the non-functional requirement interface. The object is deployed and managed on the cloud by means of the OaaS abstraction. 
Specifically, an OaaS-based platform (we call it Oparaca and introduce it in Section \ref{sec:architecture}) receives the object deployment packages from the developer,
deploys them on the cloud, and also automatically configures and monitors their resource allocation to meet the defined non-functional requirements.


The proposed paradigm greatly simplifies the process of delivering cloud-native applications. Instead of having multiple logic/data deployments with multiple rounds of development-deployment-evaluation that are subjected to many uncertainties caused by the cloud's shared environment and uncooperative abstraction realization, the application now needs to deal with only one type of abstraction. Moreover, with the non-functional requirements serving as the driving force for the underlying OaaS orchestration, no re-deployment or re-configuration is needed to meet the desired non-functional requirements.


\section{Oparaca: an OaaS Realization}
\label{sec:architecture}

In this part, we first describe the design goals of Oparaca---an open-source platform realizing the ideas of the OaaS paradigm. Then, we introduce new concepts and interfaces needed for this realization, and finally, we discuss its development details.

\subsection{Design Goals and Requirements}
\label{sec:architecture-goals-requirements}

We use Oparaca as a proof of concept to (1) illustrate how OaaS can reshape cloud application deployment, making it more productive and cost-effective; and (2) highlight how OaaS unlocks new opportunities for a more efficient, collaborative application deployment optimization. To achieve these objectives, we outline the following requirements and try to ensure Oparaca meets them throughout the entire design and implementation process.

\begin{enumerate}[leftmargin=*, noitemsep, topsep=1pt]
    \item \textit{Simplicity}: Extend the concept of \textit{object} in OOP to a service abstraction that allows application developers to encapsulate their application logic, data, and non-functional requirements into a single deployment entity.
    \item \textit{Declaratory}: Provide a simple, human-friendly interface for non-functional requirements that allows developers to express and achieve desired non-functional requirements with minimum configuration/deployment effort.
    \item \textit{Efficiency}: Oparaca can enforce application requirements at comparable cost versus state-of-the-art solutions. 
    \item \textit{Portability}: Oparaca allows applications to deliver proper functionality with desired QoS anytime, anywhere.
\end{enumerate}

Oparaca is implemented in Java and comprises approximately 20,000 lines of code. The platform offers a YAML-based OaaS API for defining objects and their non-functional requirements. Oparaca operates with FaaS functions at the container level using Knative and Kubernetes, and it provides a supported SDK for working with Python. The source code is available at \url{https://github.com/hpcclab/OaaS}.

\subsection{OaaS Abstraction Interface}
\label{sec:archtiecture-interface}

To fulfill the first two requirements (i.e., simplicity and declaratory), we provide a deployment interface for OaaS to help developers define the entities of their cloud-native application and non-functional requirements akin to OOP concepts. To that end, the cloud-native application is built on the foundation of \textit{classes}. Each class defines the structure of independent executable objects that are responsible for carrying out one or multiple functionalities. Upon deployment, Oparaca allocates appropriate cloud resources to realize the corresponding objects of the class and manage them to handle workloads. Moreover, Oparaca supports \textit{inheritance} and \textit{polymorphism} for its classes.

Within each class, we can define \textit{methods} and \textit{attributes} to encapsulate the application logic and state (that can be in the form of structured or unstructured data, i.e., BLOB), respectively. For structured state data, Oparaca allows the developer to keep the data as a JSON-based document, similar to the document database \cite{carvalho2022nosql}. For unstructured data, however, object storage is employed to store them.
We model each method as a serverless function\footnote{we use the term function and method interchangeably in this paper}. Oparaca shares object states among methods of the same object following the OOP encapsulation principles.

In Oparaca, application QoS and constraints are declared through the \textit{non-functional requirement interface}. The interface allows the developer to associate a class or its methods with one or a set of requirements that the cloud provider has to meet once objects of the assigned class or methods are deployed successfully. Table \ref{tab:non-functional-requirements} shows the list of QoS and constraints currently supported by Oparaca. Non-functional requirement declarations are treated as properties of classes or methods, so they are enforced according to the OOP inheritance principles. If a method and its class have conflicting requirements, then the method-level requirement prevails.

\begin{minipage}{0.95\linewidth}
 \linespread{0.8}
\begin{lstlisting}[
    language=yaml,
    caption=OaaS Deployment for Image Processing,
    label=lst:image-yaml
]
classes:
  - name: Image
    qos:
        availability: 99.9
    constraint:
        persistent: true
    keySpecs:
      - name: image #File Image;
    functions: 
      - name: resize
        qos:
            throughput: 100  #rps
        #container image
        image: img/resize
      - name: changeFormat
        image: img/change-format
      - name: detectObject
        qos:
            throughput: 100
        image: img/detect-object
  - name: LabelledImage
    parent: Image
    keySpecs:
      - name: labels #File labels;
    functions: 
      - name: analyze
        qos:
            throughput: 50
\end{lstlisting}
\end{minipage}
\vspace{1mm}

\begin{figure} [t]
    \centering
    \includegraphics[width=0.8\linewidth]{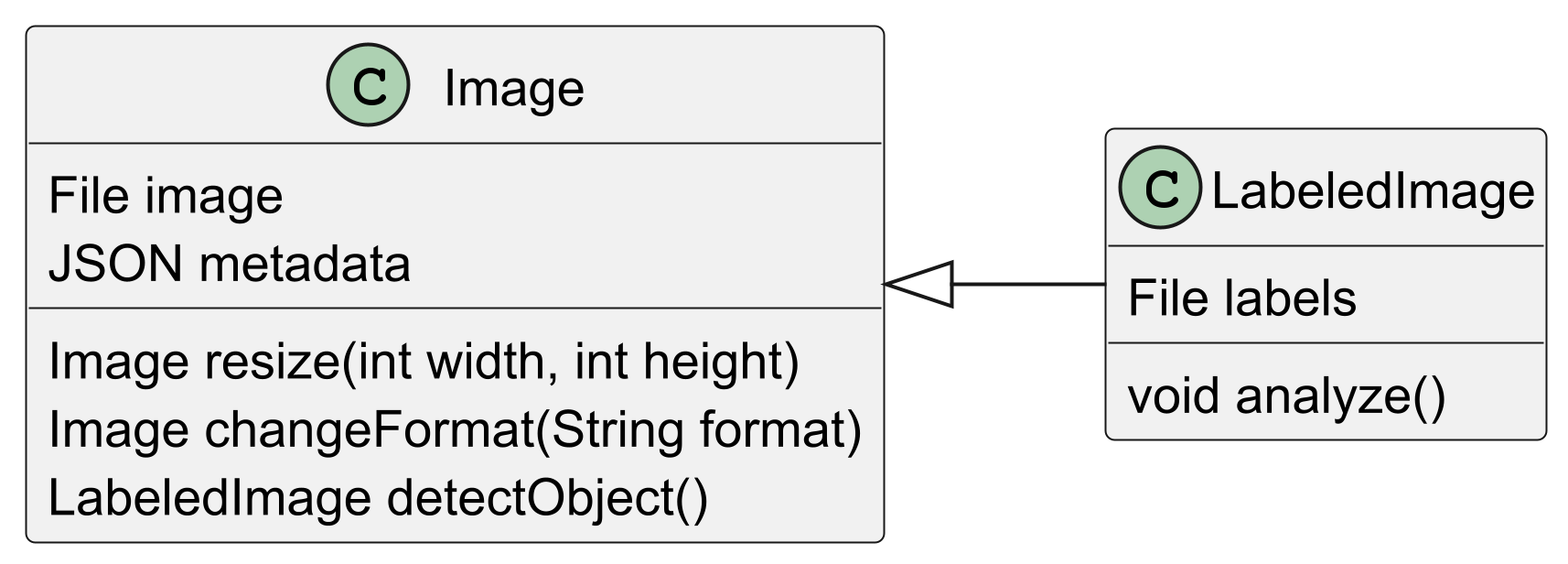}
    \caption{Class diagram for the image processing example. The developer can translate the class diagram directly to cloud deployment in Listing \ref{lst:image-yaml} through OaaS abstraction.}
    \label{fig:oaas-example}
\end{figure}

\begin{figure} [t]
    \centering
    \includegraphics[width=\linewidth]{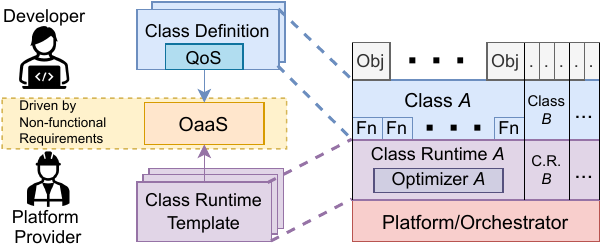}
    \caption{Realizing objects with class runtime and template: OaaS maintains templates customized for various deployment scenarios. For a specific class, Oparaca uses one of its predefined templates to create a class runtime to manage the deployed classes optimally.}
    \label{fig:class-runtime}
\end{figure}

Figure \ref{fig:oaas-example} shows the class diagram of an example application providing image processing functionalities, such as resizing and changing the format. A developer can translate the diagram directly to OaaS classes. Specifically, OaaS allows images to be wrapped inside the \texttt{Image} class abstract where the image itself can be defined as a single unstructured file and its metadata is structured data. 
The \texttt{resize} function receives width and height as its inputs and produces a new image object as its output. The \texttt{changeFormat} function receives the new format name as input and produces a new image as the output object. The developer can add a new class \texttt{LabelledImage} for the image that can have the label data of image content. This class extends the \texttt{Image} class with the additional \texttt{labels} data and \texttt{analyze} function. The \texttt{Image} class also has a \texttt{detectObject} function to perform object detection to create the \texttt{labels} data and create the \texttt{LabelledImage} object as an output. The \texttt{analyze} function is to perform further analysis to label data. Oparaca currently supports the OaaS Abstraction Interface in \texttt{YAML} format. The class declaration of the example is in Listing~\ref{lst:image-yaml}.

Based on inheritance, in this example, the \texttt{LabelledImage} class inherits the non-function parameters from \texttt{Image} class (i.e., availability=99.9). The \texttt{resize} and \texttt{changeFormat} functions that the class \texttt{LabelledImage} inherit also maintain the non-functional parameter from class \texttt{Image}. 



\subsection{Object Realization}
\label{sec:architecture-realization}



\subsubsection{Class Runtime and Template}
\label{sec:architecture-realization-cr}

Oparaca uses \textit{class runtime} to deploy and manage objects derived from user-defined classes (Figure~\ref{fig:class-runtime}).
To meet the third requirement (i.e., efficiency), the class runtime must be optimized to fulfill the non-functional requirements within a reasonable cost and overhead. However, given the non-functional requirements that Oparaca supports, there is a vast diversity of possible non-functional requirement combinations that need different specializations to satisfy. 
Thus, it is impractical to have a single design for the class runtime that can efficiently satisfy all of the requirements.

To resolve the problem, Oparaca introduces \textit{class runtime template}, which provides a configurable class runtime design optimized for a specific set of requirement combinations. 
Oparaca maintains a list of different templates to support as many requirement combinations as possible.
When deploying a class, Oparaca will choose from the list the most suitable template to realize the class requirement and then follow the template design to create a dedicated class runtime for this class. 
This approach allows Oparaca to satisfy both portability and efficiency design requirements.

In terms of portability, the class runtime template enables Oparaca to have freedom and flexibility in realizing objects. Instead of seeking a one-size-fits-all object realization mechanism, Oparaca decomposes the object realization into a set of sub-problems, each one aiming to find the optimal solution (i.e., class runtime template) for a specific infrastructure setting and requirement combinations. The approach makes Oparaca's implementation modular and flexible. One can upgrade existing solutions, extend the implementation to include new non-functional requirements, or even adjust for new infrastructure by adding/modifying templates without worrying about compatibility issues.

In terms of efficiency, Oparaca can use off-the-shelf solutions to implement its class runtime templates. This allows Oparaca to take advantage of a vast diversity of existing state-of-the-art solutions, which have been proven to be efficient in practice, to reliably enforce non-functional requirements at minimum time, cost, and effort. Further, since class runtime templates are configurable, depending on specific object deployment scenarios, the class runtime derived from the template can be customized for further efficiency. 
Oparaca also allows platform provider to customize the template configurations, selection conditions, and priority for their operation objective (e.g., resource utilization).

\begin{figure}
    \centering
    \includegraphics[width=\linewidth]{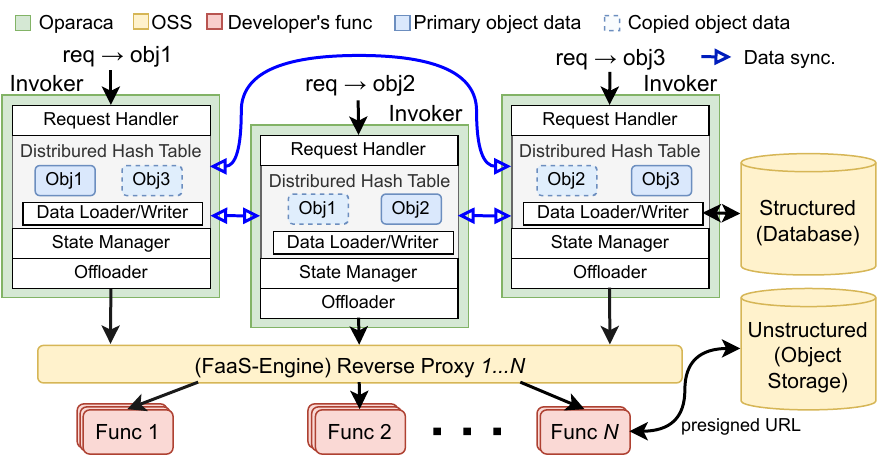}
    \caption{LTAG (Latency, Throughput, and Availability Guarantee): An example of a class runtime template designed for enforcing class latency, throughput, and availability requirements (OSS: Open-source software).}
    \label{fig:classruntime}
\end{figure}

\subsubsection{Class Runtime Example}
\label{sec:architecture-realization-ltag}

Figure \ref{fig:classruntime} shows LTAG (Latency, Throughput, and Availability Guarantee)---a class runtime template that Oparaca currently uses to enforce class latency, throughput, and availability requirements. Each class runtime derived from the template has three modules: \textit{invoker}, \textit{FaaS engine}, and \textit{data storages}. The invoker is responsible for handling all of the object-related operations. For each operation, the invoker finds its corresponding function and offloads the operation to that function managed by the FaaS engine. LTAG can maintain the object state in both unstructured and structured databases.


In the offloading mechanism, the invoker utilizes the pure function approach that bundles the invocation request and the object attributes as a standalone task within a FaaS engine. Each invocation takes the object attributes as input, modifies them, and then returns the updated attributes as the output to the invoker. The invoker maintains an internal in-memory distributed hash table (DHT) \cite{hassanzadeh2021dht} to keep the object data (i.e., attributes and metadata) for reducing database access operation, thereby speeding up the object invocation.

\vspace{2mm}
\noindent\textbf{Throughput Enforcement.}
OaaS currently supports throughput enforcement by allowing applications to specify a guaranteed invocation rate $A$ per  FaaS function \cite{nguyen2019real}. Oparaca ensures that sufficient resources are available so that at least one invocation can start immediately (i.e., without cold-start delays) every $\frac{1}{A}$ seconds. LTAG customizes the Invokers and FaaS engine based on Real-time Serverless \cite{nguyen2019real, nguyen2023storm} to estimate and periodically adjust resource allocation for each class and its functions, ensuring they can handle operation requests up to the specified rate guarantee.


\vspace{2mm}
\noindent\textbf{Latency Enforcement.}
Recent work on latency QoS aims to minimize end-to-end latency in a best-effort manner \cite{jin2023ditto, lei2023chitu, zhang2023online, zhang2021caerus, lin2020modeling}, giving no guarantee to construct/realize non-functional requirements. Besides, other efforts try to keep latency within a specific target deadline \cite{szalay2022real, ascigil2021resource, verma2024lease, moghimi2023parrotfish, bhasi2022cypress}, but this is extremely difficult from the cloud provider's perspective due to the highly dynamic and unpredictable nature of invocation logic \cite{shadrad20serverless, joosen2023does, eismann2021state}, data size \cite{bhasi2022cypress, eismann2020predicting, mvondo2021ofc}, and communication requirements \cite{yu2023following}.
Thus, to enforce the latency in a feasible and controllable way, OaaS offers guarantees to minimize the system overhead of invocation executions, focusing on cold-start and communication, enabling the developers to optimize their functionality execution time barely based on improving their codes. The developer can address cold-start via throughput enforcement, as described above. For communication, OaaS provides a \textit{locality} guarantee, allowing developers to specify the location for invocation dispatch. This can be either (i) \textit{local}: attributes are read and written as if they are in the same FaaS container executing the function logic, and (ii) \textit{none}: no locality restriction.

LTAG enforces the \textit{local} guarantee by exploiting the class function-attribute relationships. Specifically, Oparaca uses consistent hashing, maintained by invokers, to track object data locations and route invocation requests to the corresponding place.

\vspace{2mm}
\noindent\textbf{Availability Enforcement.}
OaaS provides availability enforcement as a reliability guarantee, defining the percentage of time that an object (or its methods) are available for invocation execution.
LTAG enforces availability through replication. Specifically, given an object with availability requirement $A$ (e.g., 99.99\%), we enforce $A$ by creating $N$ replicas of the object with $N$ is defined according to Meroufel and Belalem \cite{meroufel2013managing} as follows. 
\begin{equation}
    N = 1 - (1 - P)^{A}
    \label{eqn:availability}
\end{equation}
where $P$ is the stability of the resources used to deploy the object. LTAG replicates the object data and uses the DHT to manage them. However, it keeps only one object replica, called \textit{primary}, active at a time. To enforce consistency, the primary object handles all state modifications and then commits the results across all replicas. If the primary replica fails, Oparaca chooses one of the remaining replicas as the new primary. 

\subsection{Oparaca Architecture}
\label{sec:oaas-architecture}

\begin{figure}
    \centering
    \includegraphics[width=\linewidth]{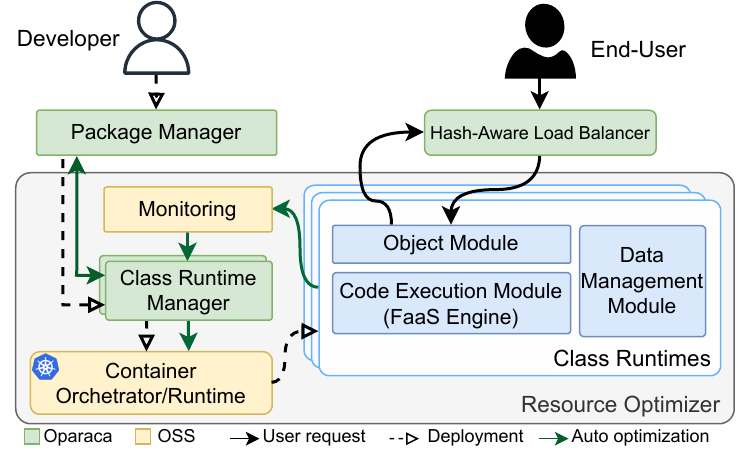}
    \vspace{-2mm}
    \caption{A bird-eye view of Oparaca's architecture}
    \label{fig:overall-architecture}
\end{figure}

Oparaca's architecture, shown in Figure \ref{fig:overall-architecture}, includes the following key components: \textit{(1) Package Manager}: responsible for managing classes registered in Operaca and their corresponding deployment packages. This component also acts as a gateway and offers APIs to develop and deploy OaaS-based applications. \textit{(2) Class Runtime}:    turns the class descriptions and corresponding packages into the actual object deployments on the cloud. \textit{(3) Class Runtime Manager}: create dynamic class runtime from existing templates (e.g., LTAG). It is also responsible for class runtime deployment and management. \textit{(4) Monitoring System}: gathers the performance metrics from class runtime. \textit{(5) Hash-aware Load Balancer} and \textit{Container Runtime}: responsible for scheduling and managing function execution. Once a function invocation is issued, the hash-aware load balancer routes the request to the corresponding class runtime by using consistent hashing that, in turn, forwards the request to the corresponding container for execution. 

Given the interface and architecture, the application lifetime on the cloud now consists of two phases: 

\textbf{(a) Registration:} The developer registers their class to Oparaca. Upon registration, the \textit{package manager} unpacks the deployment, extracting the class logic (e.g., functions), state (e.g., data schema), and non-functional requirements (e.g., QoS and constraints). The extracted information is then forwarded to the \textit{class runtime manager} to find an appropriate class runtime template to generate a dedicated \textit{class runtime} to handle the object realization for the class.

\textbf{(b) Execution:} Once a \textit{class runtime} is created, it is responsible for managing the execution and state of all objects generated from the associated class. Every interaction with the application users is handled through the class runtime, independent from other Oparaca components. To ensure reliability, the \textit{class runtime manager} periodically collects monitoring metrics from class runtime. Based on the information, Oparaca can adjust the \textit{Container Orchestrator/Runtime} to improve efficiency and take administrative actions (e.g., to recover from failure, etc.) if needed.

Note that the above procedures are performed solely by Oparaca platform. Application developers do not have to intervene or refine their configuration for both functional and non-functional requirements. This greatly simplifies application deployment.

\section{Evaluation}
\label{sec:evaluation}

In this section, we seek to learn the performance of Oparaca in the following aspects: 
non-functional requirement enforcement (Section~\ref{sec:evaluation-non-functional}), 
implementation efficiency (Section \ref{sec:evaluation-implementation}), 
deployment productivity (Section \ref{sec:evlt:deployment}), and development productivity (Section \ref{sec:evaluation-ease-of-use}).

\subsection{Experimental Setup}
\label{sec:evaluation-setup}

\begin{figure} [t]
    \centering
    \includegraphics[width=0.9\columnwidth]{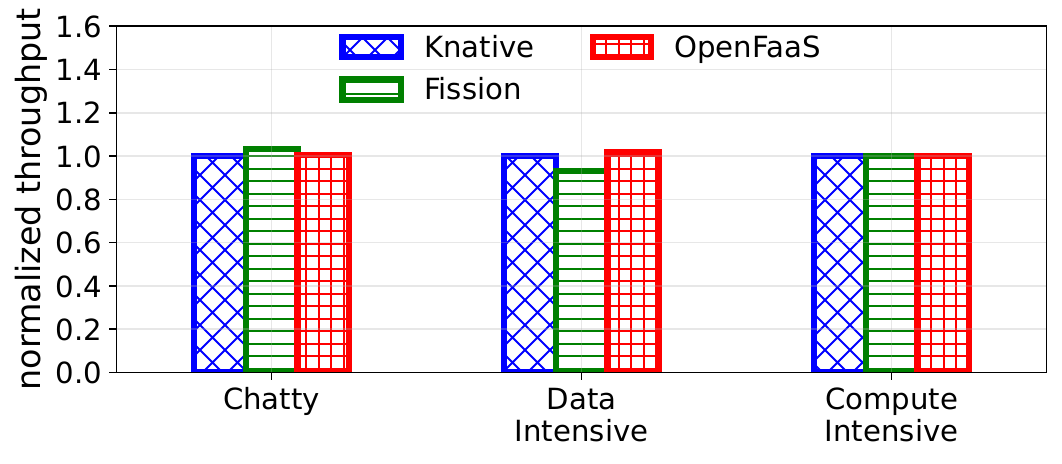}
    \caption{Oparaca does not significantly differ in throughput performance across the FaaS engines.}
    \label{fig:faas-engines}
\end{figure}

\begin{figure*}[ht]
    \centering
    \subfloat[Chatty]{\includegraphics[width=0.33\textwidth]{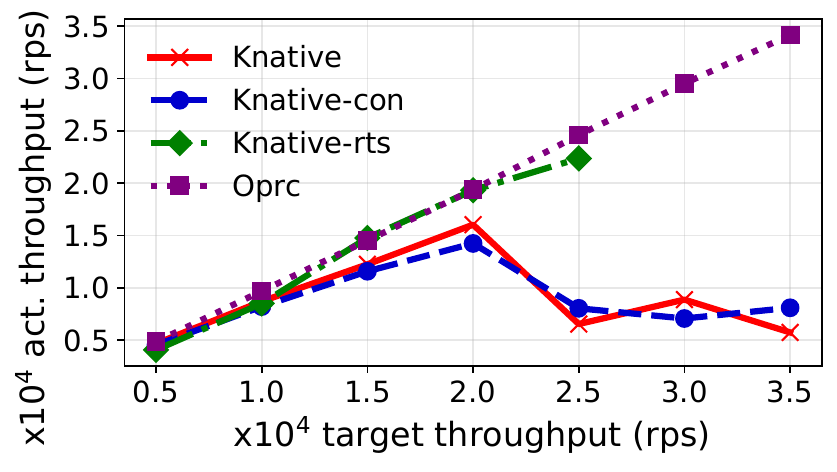}\label{fig:evlt:tp:json}} 
    \hfill
    \subfloat[Data Intensive]{\includegraphics[width=0.33\textwidth]{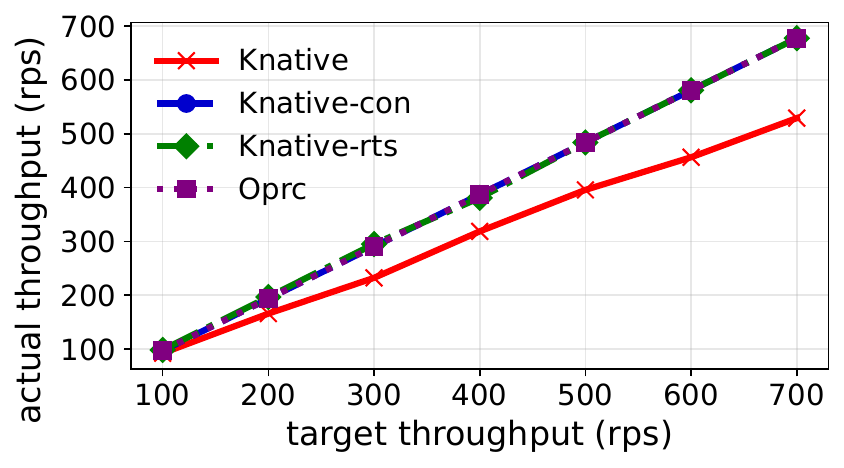}\label{fig:evlt:tp:image}}
    \hfill 
    \subfloat[Compute Intensive]{\includegraphics[width=0.33\textwidth]{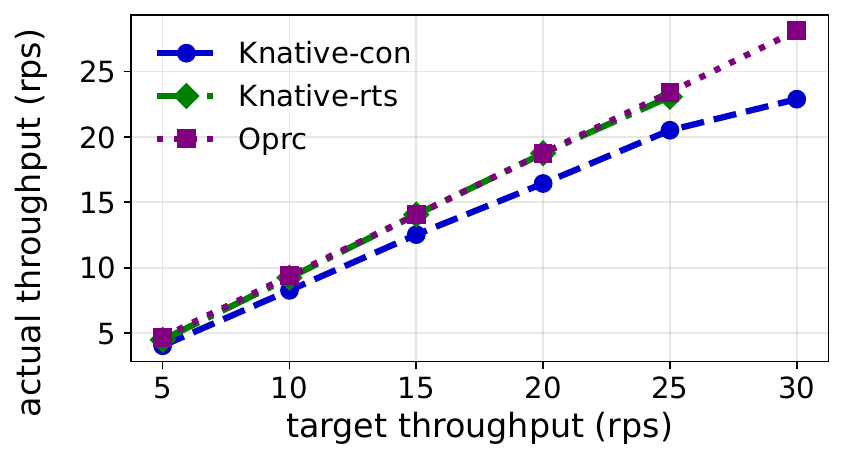}\label{fig:evlt:tp:video}}

    \caption{Achievable throughput varying target throughput. Oparaca ensures the actual throughput matches the target one across settings, while the other approaches fail to do so at high throughput targets.
    }
    \label{fig:evlt:tp-enforcement}
\end{figure*}

\begin{figure}
    \centering
    \includegraphics[width=\columnwidth]{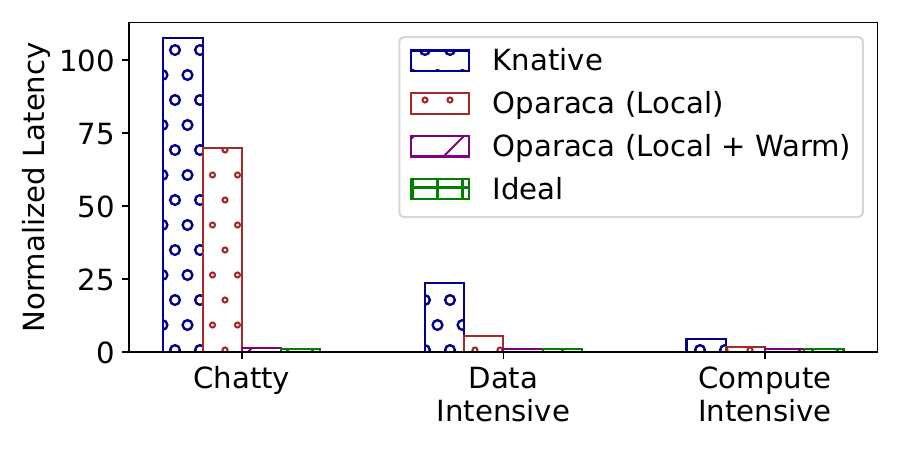}
    \caption{Oparaca can exploit data locality to provide various latency guarantees.}
    \label{fig:latency-enforcement}
\end{figure}

\begin{figure} [ht]
    \centering
    \includegraphics[width=\linewidth]{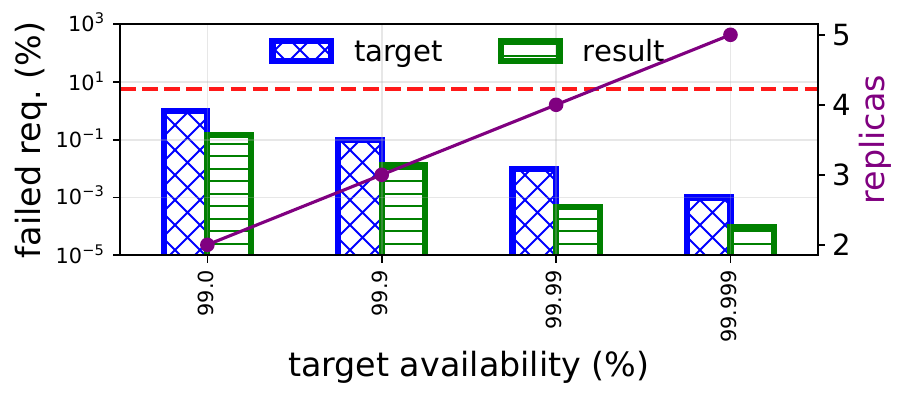}
    \caption{
    Successful invocation rate at different availability targets with availability enforcement. Resource stability ($P$) is 94.36\% (red line).
    }
    \label{fig:evlt:availability}
\end{figure}

We prepare the experimental environment on 4 machines on Chameleon Cloud \cite{chameleon_cloud}, each with 2 sockets of Intel(R) Xeon(R) Gold 6240R CPU processors that collectively have 192 cores, 768 GB memory, and SSD SATA storage. We use 3 machines to install the Kubernetes cluster (RKE2~\cite{rke2}) for deploying applications. The last machine generates load using Gatling~\cite{gatling}. Regarding data management, we use Minio~\cite{minio} (S3-compatible storage) for unstructured data and  ArangoDB~\cite{arangodb} (document database) for structured data.

\vspace{1mm}
\noindent\textbf{Workloads.} To make sure our evaluation is comprehensive, we consider the following three classes of applications that exhibit different behaviors: 
\begin{itemize}[leftmargin=*, noitemsep, topsep=0pt]
    \item \textit{Chatty}: characterized by frequent small communications that impose significant overhead on network transmission \cite{chatty-app}. As a representative workload for the application class, we utilize \textit{JSON randomization} \cite{lloyd2018serverless}, which involves a sequence of ten invocation requests, each randomly updates a JSON key-value pair to the document database. 
    \item \textit{Data Intensive}: characterized by substantial data access operations \cite{hoseinyfarahabady2021data}. We use an \textit{image resizing} workload \cite{solaiman2020wlec, balla2021estimating}, which resizes images stored in object storage through FaaS invocations, to represent this class of applications
    \item \textit{Compute Intensive}: demand extensive computational resources throughout their lifecycle (e.g., ML \cite{chahal2020migrating} and HPC \cite{nguyen2020motivating} applications). To represent this class, we use \textit{video transcoding} \cite{moina2023event, wu2020descriptive}, which involves changing the resolution of a video file stored in object storage.
\end{itemize}



\vspace{1mm}
\noindent\textbf{Approaches.}
To ensure generality, we integrated Oparaca with various FaaS engines---Knative~\cite{knative}, Fission~\cite{fission}, and OpenFaaS~\cite{openfaas}, all backed by Kubernetes---to host object functions. Figure \ref{fig:faas-engines} shows the maximum throughput achieved by workloads mentioned above when deployed over Oparaca using these different FaaS backends under identical resource configurations (each deployment can scale up to five Kubernetes pods, each with 4 CPUs). The throughputs, normalized to Knative, are nearly equivalent across all FaaS engines for all three workloads. This confirms that Oparaca can be configured to work with various FaaS engines with negligible performance differences, making it flexible for deployment across different cloud environments.
Thus, due to space limits, we report only the experimental results for Oparaca's Knative variant. Also, for fair comparison, we use Knative with various deployment configurations as experiment baselines:
\begin{itemize}[leftmargin=*, noitemsep, topsep=0pt]
    \item \textit{Knative:} Default Knative configuration that declares only per-container resource requirements (i.e., CPU and memory) and leaves the rest to the auto-scaling system.
    \item \textit{Knative-con:} Default Knative configurations plus applying per-container concurrency limit to avoid overloading.
    \item \textit{Knative-rts:} adopt Real-time Serverless resource management \cite{nguyen2019real} to enforce throughput guarantee.
    \item \textit{Oprc} is Oparaca, which allows the applications to enforce their throughput, latency, and availability in their class definitions. Since Oparaca needs to learn the workload metrics before properly optimizing the class runtime, we perform one more extra round of load generating in each experiment. The first round acts as the warm-up for Oparaca to properly gather the metrics. 
\end{itemize} 
Beyond ensuring a fair comparison, we choose Knative as a FaaS baseline because it offers a rich set of configuration options to capture diverse deployment scenarios often unsupported by other engines. Additionally, varying Knative settings demonstrate how current FaaS implementations address non-functional requirements---by adjusting low-level configurations (e.g., per-container concurrency) in a best-effort manner. Configuring Knative allows us to explore a broad range of FaaS deployment configurations, whether these adjustments are made by developers (if the FaaS engine exposes the configurations) or by cloud providers (if it does not---for example, Microsoft Azure doesn't allow developers to configure per-container concurrency). Thus, although our evaluation results are specific to Knative, the insights and implications are generalizable to other FaaS engines.

\begin{figure*}[ht]
    \centering
    \subfloat[Chatty]{\includegraphics[width=0.33\textwidth]{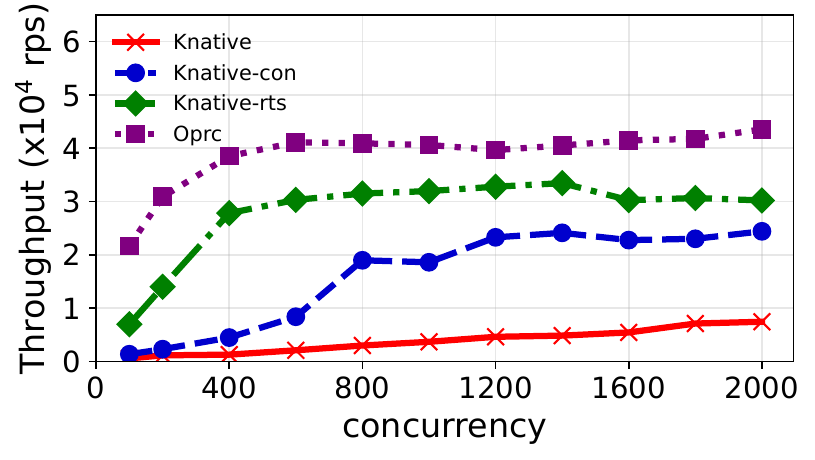}\label{fig:evlt:scalability:json}} 
    \hfill
    \subfloat[Data Intensive]{\includegraphics[width=0.33\textwidth]{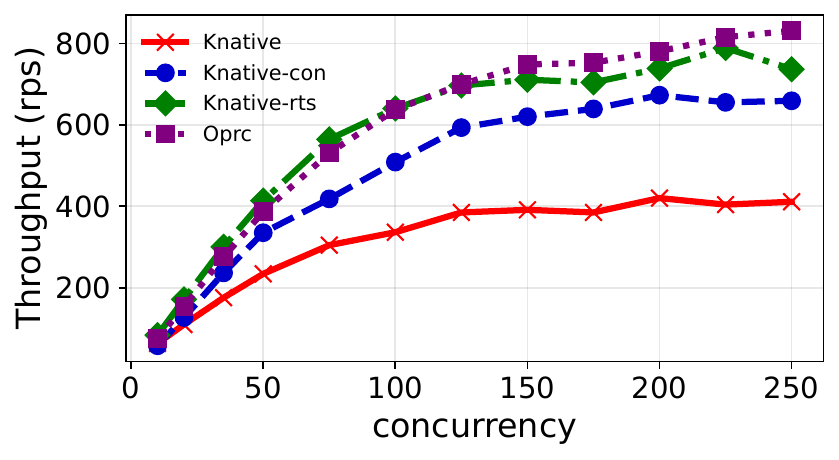}\label{fig:evlt:scalability:image}}
    \hfill 
    \subfloat[Compute Intensive]{\includegraphics[width=0.33\textwidth]{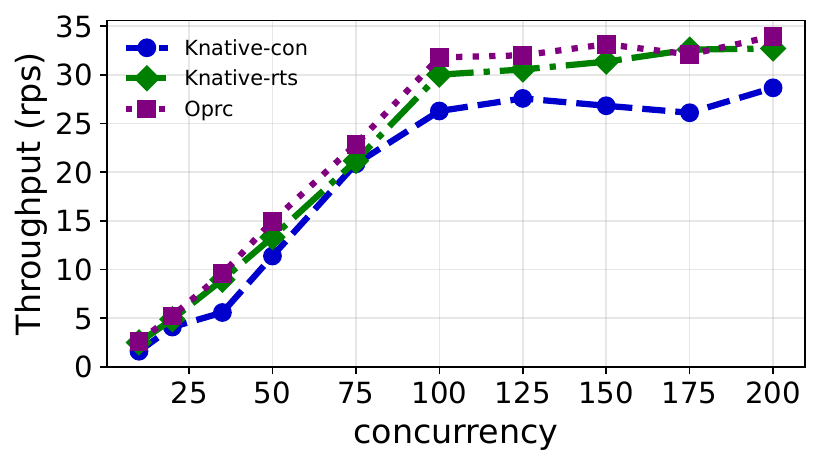}\label{fig:evlt:scalability:video}}

    \vspace{-2mm}
    \caption{
    Achievable throughput under various request concurrency.
    Concurrency is defined as the number of clients that concurrently generate requests for the system.
    }
    \label{fig:evlt:scalability}
    
\end{figure*}

\begin{figure*}[ht]
    \centering
    \subfloat[Chatty]{\includegraphics[width=0.33\textwidth]{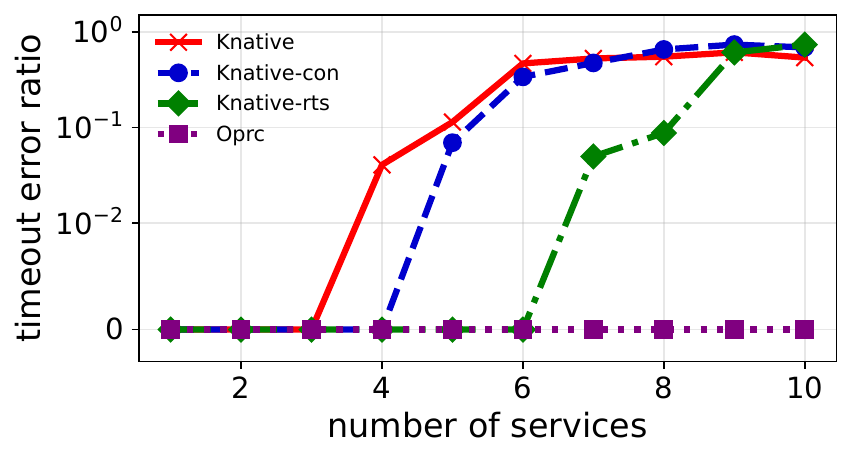}\label{fig:evlt:efficiency:json}} 
    \hfill
    \subfloat[Data Intensive]{\includegraphics[width=0.33\textwidth]{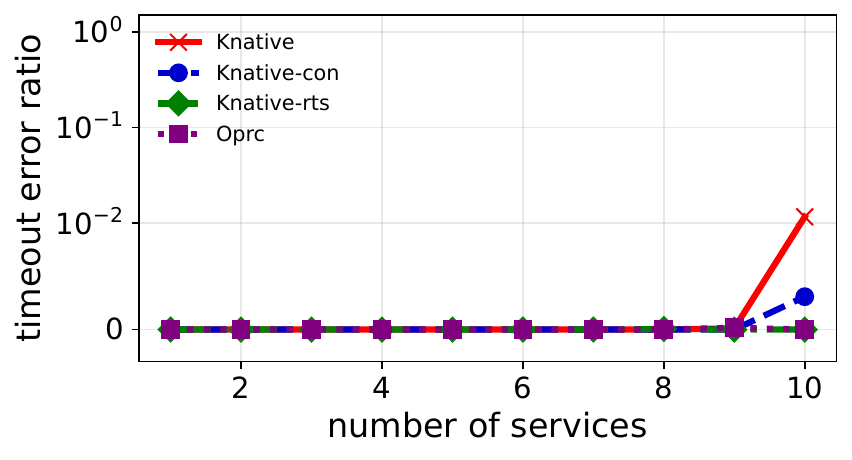}\label{fig:evlt:efficiency:image}}
    \hfill 
    \subfloat[Compute Intensive]{\includegraphics[width=0.33\textwidth]{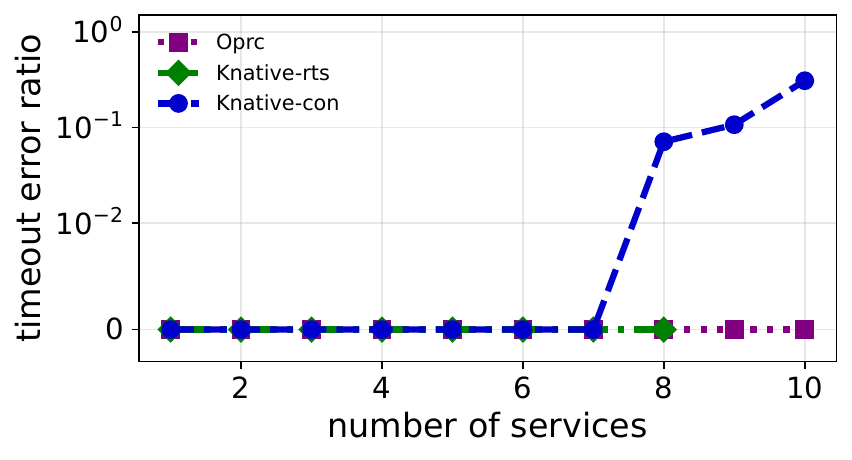}\label{fig:evlt:efficiency:video}}

    \vspace{-2mm}
    \caption{Error response ratio of different solutions upon deploying them with the different number of services.
    }
    \label{fig:evlt:efficiency}

\end{figure*}

In the following experiments, Oparaca deploys and manages workloads using class runtime derived from the LTAG template. Thus, data access is automated via the invoker.
In the Knative variants, however, these applications have to implement direct data access to storage or database manually.

\subsection{Experimental Results}

\subsubsection{Non-functional Requirement Enforcement}
\label{sec:evaluation-non-functional}

We validate the QoS enforcement capability of Oparaca by deploying applications mentioned in Section \ref{sec:evaluation-setup} using the LTAG class runtime template as described in Section \ref{sec:architecture-realization-ltag}.


\vspace{2mm}
\noindent\textbf{Throughput.}
To validate Oparaca's throughput enforcement, we deployed the three applications with various target throughputs. Then, we configured the load generator to send the request at the same rate as the target throughput and measured the actual throughput on each system. The results are reported in Figure~\ref{fig:evlt:tp-enforcement}.

Overall, Oparaca can guarantee the throughput for all three applications. \textit{Knative-rts} only meets low throughput targets and fails at higher ones due to over-provisioning.
The other two \textit{Knative} variances fail to meet the targets since they only rely on auto-scaling without the awareness of the target throughput.
In the chatty workload, with the high request arrival rate, the internal queue cannot hold requests long enough to wait for the new pod to be spawned. Meanwhile, in the compute-intensive application, it takes longer for each request to be processed, making it easier to time out. Only the data-intensive application that \textit{Knative-con} can meet the target throughput.

The results also demonstrate the complexity of FaaS configuration. Even when utilizing the same backend services (i.e., Knative), varying FaaS deployment configurations result in significantly different performance outcomes. Thus, manual adjustment of FaaS deployment, while daunting, is often required to achieve the desired throughput. In contrast, Oparaca simplifies and automates this process with its high-level interface.  




\vspace{2mm}
\noindent\textbf{Latency.} We deployed all three applications over Operaca under the \textit{locality} and \textit{throughput} guarantee. We let the applications run under bursty loads by configuring the load generator to remain idle most of the time but occasionally create sudden bursts that send requests at a rate equal to the application throughput guarantee for a very short duration.
We compare Oparaca against two baselines: (i) \textit{Knative} with the data storage deployed at a separate data center from the FaaS deployment, representing a typical scenario of FaaS deployment \cite{aws-security-resilient}, and (ii) \textit{Ideal} where functions and data storage are deployed together on a dedicated machine with excessive resources, representing an ideal execution environment where the invocation execution latency depends solely on the application itself. 

Figure \ref{fig:latency-enforcement} shows the average execution time of the three applications across different deployments. The latency is normalized to the case of the ideal deployment. Knative is the worst among approaches, with the latency can be as high as 60$\times$ the ideal. The reason is two-fold. First, Knative needs external storage to keep the application data, but the actual data location is hidden under the storage abstraction, causing significant data transmission latency. Oparaca does not have this limitation as it encapsulates the data and invocations under a unified object abstraction, enabling locality enforcement, i.e., \textit{Oparaca (Local)}, that allows invocations to execute at the same machine with their data, significantly reducing the latency by 1.5$\times$ (Chatty) to 4$\times$ (Data Intensive). Second, Knative scales resources allocated to FaaS functions based on concurrency. That makes invocations suffer from cold-start under bursty loads. Applications can workaround this issue with Oparaca via throughput guarantee, enforcing the cloud to execute invocations without cold-start up to a certain rate, i.e., \textit{Oparaca (Local + Warm)}. This configuration further reduces the latency by 1.7$\times$ (Compute Intensive) to 46.5$\times$ (Chatty)! Enforcing these two non-functional requirements together allows applications deployed over Oparaca to minimize their invocation overhead (as low as 7\% of execution time), achieving invocation execution latency that is very close to the ideal execution.

\vspace{2mm}
\noindent\textbf{Availability.}
Next, we validate Oparaca's availability enforcement. We have created a failure emulator that injects failures by deleting the platform container according to a predefined Mean Time Between Failures (MTBF). Whenever a failure is injected, Kubernetes automatically recovers the container. The emulator then waits for MTBF, which is also supplemented by a random value from a normal distribution, before introducing the next failure. The emulator carries out these operations on each container individually. 
To select the MTBF, we use the reference MTBF of the Intel server boards~\cite{intel-board} that have around 50K hours on average. To speed up the experiment process, we scaled this number down by a million, setting the MTBF to 180s, which makes each container only operate for 94.36\% of the time. We then use 94.36\% as the resource stability ($P$) to configure Oparaca. We deploy the application according to the different target availability, generate the load to test the actual application availability with a rate of 200 requests per second for 1.5 hours, and measure the ratio of the requests being processed unsuccessfully. 

The results of this experiment are reported in Figure \ref{fig:evlt:availability}. 
When availability enforcement is on, Oparara deploys classes and objects with replications, significantly reducing the failure rate to meet the availability targets. The actual failed request ratio is slightly lower than each predefined target because Oparaca adds just enough replicas to meet the target, minimizing availability enforcement overhead. Notably, increasing the availability from 99\% to an exceptional rate of 99.999\% (1000$\times$ better) incurs only 2.5$\times$ extra resource cost. This is a 50$\times$ improvement versus the current industry standard that necessitates an SLA on availability of 99.95\% \cite{aws-lambda-sla} with only 1.67$\times$ cost increment.




\vspace{2mm}
\noindent
\colorbox{blue!10}{
\parbox{0.96\linewidth}{
\underline{\textbf{Takeaway}:} \emph{
Unlike traditional FaaS deployments, Oparaca can automatically reconfigure to enforce various non-functional requirements for different classes of applications, eliminating the need for manual refinement.   
}}}

\subsubsection{Efficiency of Oparaca}
\label{sec:evaluation-implementation}

In this subsection, we examine Oparaca efficiency, running various experiments on a fixed quantity of resources to see how well the implementation handles various workloads under different operation scenarios.



\vspace{1mm}
\noindent \textbf{Function Invocation Efficiency.}
To evaluate Oparaca invocation efficiency, we compare its maximum throughput with Knative variants; all are under limited resources. The throughput measurement takes multiple runs with an increasing number of clients (i.e., concurrency). We measure the mean throughput achieved in each run and report them in Figure \ref{fig:evlt:scalability}.

In general, the throughput becomes steady after increasing the concurrency to a certain level. Oparaca provides a higher throughput compared to other baselines, especially for the chatty workload (Figure~\ref{fig:evlt:scalability:json}) because Oparaca relies on the internal in-memory distributed hash table (DHT) to store the object data; thereby, it speeds up the data access and reduces the database operation. For the chatty workload, \textit{Knative-con} and \textit{Knative} yield significantly lower throughput compared to \textit{Knative-rts}. This is because this workload performs little computation compared to its network I/O operation, which makes the Knative auto-scaler inaccurately adapt the acquired resources to the workload.


For the data-intensive workload (Figure~\ref{fig:evlt:scalability:image}), \textit{Knative} performs poorly because the auto-scaler cannot accurately adjust acquired resources to the increasing workload without per-container concurrency declaration. In contrast, by only declaring per-container concurrency, \textit{Knative-con} can perform with a little less performance than \textit{Knative-rts}.

For the compute-intensive workload (Figure~\ref{fig:evlt:scalability:video}), because it is computationally intensive and the invocation rate is also less than the other workloads, all of the solutions can provide similar performance. Only \textit{Knative} cannot be used for this workload because without controlling the concurrency, each function container has to handle more concurrent invocations than it can. As a result, they fail to handle requests continually. Oparaca can perform slightly better than the others because it eliminates the need to fetch and deserialize the record (i.e., metadata) from the database on each function container.




\vspace{1mm}
\noindent \textbf{Throughput Enforcement Efficiency.}
Our primary objective in this experiment is to examine the resource efficiency of Oparaca against other baselines and ensure its throughput is not attained with the cost of lavishly allocating resources. The other objective is to investigate Oparaca's behavior in the face of services with different throughput expectations. To that end, we configure multiple services of the same type, each with its own target throughput. To achieve this, we started by testing on a single service and gradually increasing the number of services to ten. We set the target throughput of each replicated service to be 1/10th of the maximum throughput we found in the previous experiment. We chose these numbers so that the target throughput is not too low and scaling remains relevant. The experiment is performed by generating invocation requests to each service, with the request rate capped to the target throughput, and then measuring the ratio of the number of timeout errors to the total number of requests. 

As shown in Figures~\ref{fig:evlt:efficiency}, overall, Oparaca outperforms other baselines for almost all workloads. For the chatty workload (Figure~\ref{fig:evlt:efficiency:json}), Oparaca can handle all of the requests with zero error rate because of its ability to readjust its allocated resources and its internal DHT structure. \textit{Knative-rts} also performs well at the beginning; however, after 6 services, the external document database starts to slow down, leading to a sharp increase in the error rate. The poorer performance of \textit{Knative} and \textit{Knative-con} is mainly because their independent scaling of services and lack of awareness of performance objectives lead to resource contention among co-existing services.

For the data-intensive workload (Figure~\ref{fig:evlt:efficiency:image}), all baselines are capable of handling requests up to 9 services. Nonetheless, for 10 services, only \textit{Knative-rts} and Oparaca remain error-free. \textit{Knative-con} and \textit{Knative} still suffer from the resource contention. Similarly, for compute-intensive workload (Figure~\ref{fig:evlt:efficiency:video}), \textit{Knative-rts} and \textit{Knative-con} only have enough resources to meet the target throughput up to 8 and 7 services without any error, respectively. Oparaca, however, can handle all of the requests for up to 10 services.




\vspace{2mm}
\noindent
\colorbox{blue!10}{
\parbox{0.96\linewidth}{
\underline{\textbf{Takeaway}:} \emph{
Being cognizant of performance objectives is crucial for Oparaca to deliver competitive efficiency for both the user and the system across different applications while also offering a high-level abstraction to the user.
}}}

\subsubsection{Deployment Productivity Using Oparaca}
\label{sec:evlt:deployment}

\begin{figure} [ht]
    \centering
    \includegraphics[width=0.95\linewidth]{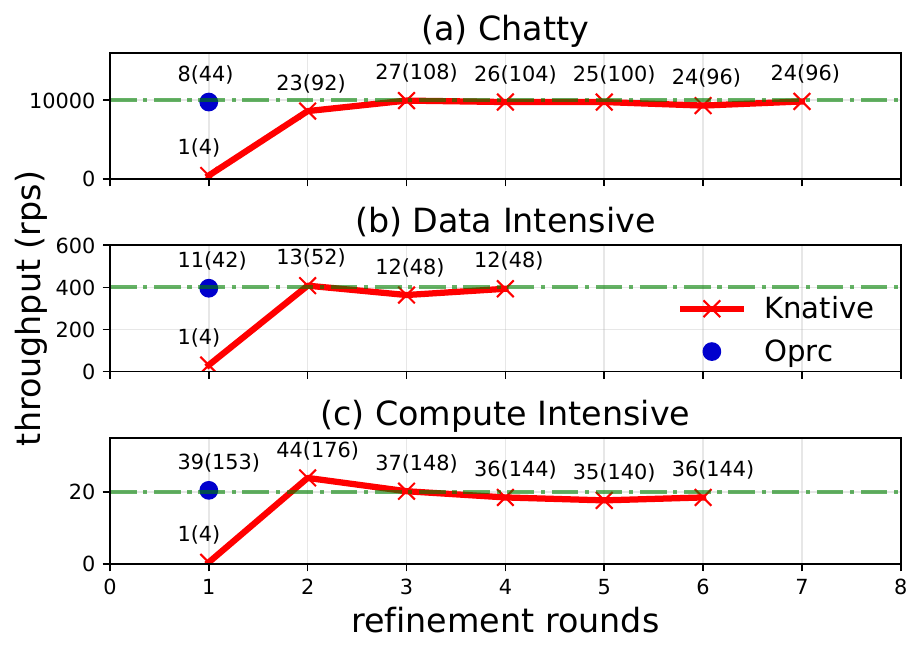}
    \caption{Rounds of refinement for Knative to enforce the target throughput (green lines) versus Oparaca. Data points are annotated by \texttt{\#pods(\#cores)}, including invoker pods.}
    \label{fig:evlt:deployment}
\end{figure}

\begin{figure} [h]
    \centering
    \includegraphics[width=1\linewidth]{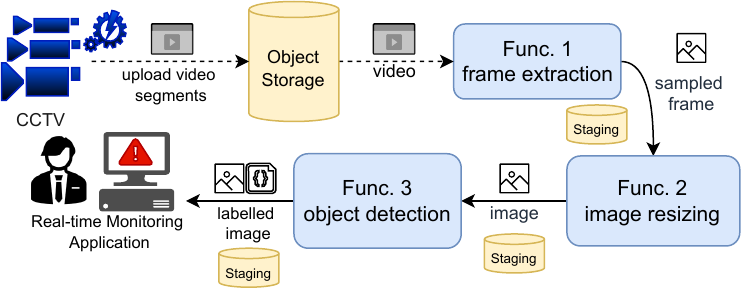}
    \caption{The case study of developing video and image processing for a real-time monitoring system}
    \label{fig:evlt:case-study}
\end{figure}

To show the productivity improvement of Serverless application deployment, we present the experiment on the refinement steps using Knative on three application deployments with the requirement to enforce the throughput of 10k, 400, and 20 requests per second for chatty, data-intensive, and compute-intensive, respectively. The manual refinement strategy consists of three phases. First, we want to find the number of pods that roughly provide throughput that is equal to our objective. We deploy the application with a single pod and then perform load testing to find the throughput. Then, we scale it up using the formula below and repeat this process until the throughput matches the objective. 
\[ pods_{next} = \dfrac{throughput_{target}}{throughput_{current}}\times{pods_{current}} \]
The second phase reduces the pods until they cannot satisfy the target. The last phase increases the container-level concurrency but reduces the number of pods to improve utilization. 

As shown in Figure~\ref{fig:evlt:deployment}, the manual refinement method needs at least 4 rounds to find the optimal number of pods to meet the target throughput, while we only need to give the Oparaca the number, and it will automatically adjust the deployment when we feed the load. Furthermore, Oparaca improves application performance while reducing the required resource allocation to meet the target throughput. For IO-intensive workloads focused on structured data like the chatty workload, Oparaca reduces resource usage from 100 cores to 44 cores. This is because OaaS unlocks cross-domain optimization---in this case, data locality---to speed up invocation execution time, quickly freeing up FaaS pods for higher concurrency and significantly reducing resource requirements compared to Knative.
Even for the compute-intensive application, where locality is not an issue, Oparaca automatic refinement still achieves the throughput target at a comparable cost (153 cores) versus Knative (144 cores,  only 6\% higher), which requires much more effort in manual tuning (6 rounds of refinement versus one).

\vspace{2mm}
\noindent
\colorbox{blue!10}{
\parbox{0.96\linewidth}{
\underline{\textbf{Takeaway}:} \emph{
Oparaca's OaaS abstraction improves deployment productivity and performance enforcement effectiveness.
}}}

\subsubsection{Development Productivity Using Oparaca}
\label{sec:evaluation-ease-of-use}

In this part, we provide two cloud application developments representing common cloud applications at different scales, non-functional requirements, and complexities. We will deploy these applications using the OaaS paradigm and recommended FaaS deployment practices to demonstrate how OaaS can make the development of cloud-native serverless applications more productive.

\vspace{2mm}
\noindent\textbf{\underline{Case Study \# 1. Real-time Monitoring System.}}
Figure \ref{fig:evlt:case-study} shows a CCTV system uploading video segments to object storage, waiting to be processed by a workflow of function that includes \texttt{extractFrame()} that splits a video segment into multiple frames; \texttt{resizeImg()} whose job is to resize the image frame to be usable by the next function in the pipeline; and \texttt{detectObject()} is in charge of performing the object detection on an image and generating label in the \texttt{JSON} format. These functions must persist their output data so that the following function in the workflow can consume it. Because the entire workflow is latency sensitive, the execution rate of the whole workflow (i.e., throughput) has to be guaranteed. Developers can calculate the throughput by the number of cameras and the object detection frequency. 

\noindent\textbf{FaaS implementation.}
The developer must repeat the following steps for each function deployment: (i) Configuring cloud-based object storage, database and maintaining the credential access token for the functions to use. (ii) Implementing the functions' business logic. (iii) Data management within the functions that itself involves three steps: (a) allocating the storage addresses to fetch or upload data; (b) authenticating access to the object storage via the access token; and (c) implementing the fetch and upload operations on the allocated addresses.
Upon implementing these functions, the developer must connect them as a workflow via a function orchestrator service (e.g., AWS Step Functions~\cite{aws-sf}). Finally, upon arrival of a new video segment, the event triggers the workflow to put the result into the database, waiting to be processed by the monitoring system. To ensure the target throughput, developers have to go through multiple rounds of testing and refinement to get the final configuration for each function.

\noindent\textbf{OaaS implementation.} The developer defines three classes: 
\begin{itemize}[leftmargin=*, noitemsep, topsep=0.5pt]

\item \texttt{\textbf{Video}} class with \texttt{extractFrame()} function that produces \texttt{LabeledImage} as the output, and \texttt{wfDetectObject(freq)} workflow function that has a detection frequency as the input. This class also has \texttt{video} file as an unstructured state.
    
\item \texttt{\textbf{Image}} class contains \texttt{resize} function and \texttt{image} file as an unstructured state (see Listing~\ref{lst:image-yaml}). 
    
\item \texttt{\textbf{LabeledImage}} class inherits from the \texttt{Image} class and has its own \texttt{objectDetection()} function and \texttt{labels} data (state) in JSON format (see Listing~\ref{lst:image-yaml}).

\end{itemize}


Upon uploading a new video to the Oparaca platform by the CCTV system, it creates a ``video'' object and invokes \texttt{video}. \texttt{wfDetectObject(freq)} that outputs a \texttt{LabeledImage} object that is consumed by the real-time monitoring application. We note that, in developing the class functions, the developer does not need to be involved in the data locating and authentication steps. To ensure the application performance, developers only need to declare the target throughput within the class definition (see example in Listing~\ref{lst:image-yaml}); then, the Oparaca can transparently create the suitable class runtimes and their configuration.

\begin{figure}[t]
    \centering
    \subfloat[FaaS-based \cite{aws-searchable-document-repo}]{
        \includegraphics[width=1\columnwidth]{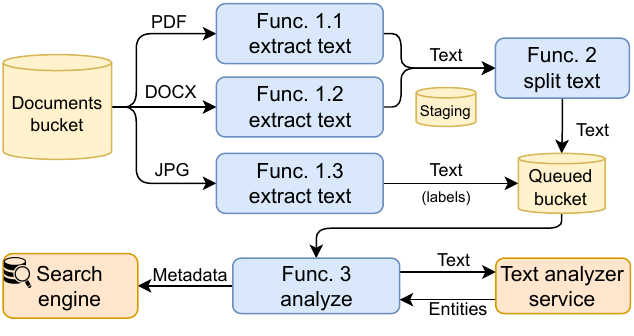}
        \label{fig:evlt:case-study2-faas}
    }
    
    \subfloat[OaaS-based]{
        \includegraphics[width=1\columnwidth]{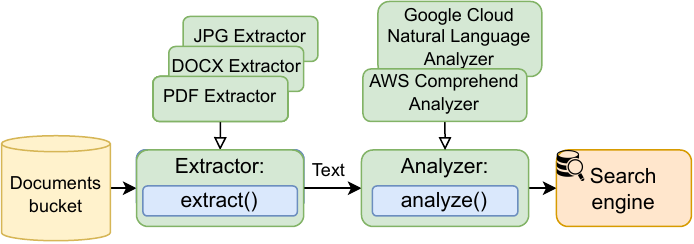}
        \label{fig:evlt:case-study2-oaas}
    }
    \caption{The searchable enterprise document repository implemented based on FaaS and OaaS paradigm.}
    \label{fig:evlt:case-study2}
\end{figure}



\vspace{2mm}
\noindent\textbf{\underline{Case Study \# 2. Searchable Document Repository.}}
Retrieving and processing at scale the vast repositories of valuable documents, images, and media from enterprise customers is a common practice in the cloud \cite{nguyen2020motivating, wang2018efficient}. In this case study, we first present how the application is deployed with traditional FaaS on the cloud, the limitations of this approach, and how to resolve them with OaaS/Oparaca. 

\vspace{1mm}
\noindent\textbf{FaaS implementation.}
Figure~\ref{fig:evlt:case-study2} shows the serverless workflow to analyze the document in various formats and update the metadata to the search engine recommended by AWS \cite{aws-searchable-document-repo}. Upon the document uploads to the document bucket (object storage), the storage triggers the event to invoke \texttt{extractText()} based on the type of the document. If the document is in \texttt{PDF} or \texttt{DOCX} format, the function extracts the text and sends the text to be split by the next function \texttt{splitText()}. The result will be put into the \texttt{Queued bucket}. Alternatively, if the document is in \texttt{JPG} format, the \texttt{extractText()} function analyzes the image to get labels and puts them in the \texttt{Queued bucket}. In the next step, the \texttt{analyze()} function loads text from the \texttt{Queued bucket} to analyze it via the external text analyzer service (e.g., AWS Comprehend) and then saves the metadata result to the search engine.


The FaaS implementation has two main drawbacks. First, developers must explicitly manage application state and data using separate storage services, which increases complexity and makes it difficult to configure non-functional requirements as in the previous case study. Second, functionalities may require numerous and heterogeneous FaaS deployments—for example, needing separate extraction functions for each document type, where some (like \texttt{PDF} and \texttt{DOCX}) require staging and others (like \texttt{JPG}) do not. These drawbacks complicate development, deployment, and management as the application evolves to handle various document types and integrates more functionalities and options (e.g., using multiple text analyzer services instead of one).

\noindent\textbf{OaaS implementation.} To demonstrate the feasibility of OaaS in production, we transform the given FaaS-based solution into OaaS with minimal effort to resolve the previously mentioned drawbacks. The transformation involves three steps.

\begin{itemize}[leftmargin=*, noitemsep, topsep=0.5pt]
    \item \textbf{Workflow Construction.} We encapsulate related FaaS functions, states, and key data into objects representing two key functionalities: \texttt{\textbf{Extractor}} to extract text from the document repository and \texttt{\textbf{Analyzer}} to analyze the extracted text. The two classes form the critical path of the application processing pipeline, as shown in Figure \ref{fig:evlt:case-study2-oaas}.
    \item \textbf{Object Encapsulation.} We apply inheritance and polymorphism to promote software reuse by wrapping corresponding FaaS functions and states into classes derived from the two base classes. This approach hides the need for storage services behind the object abstraction and outsources their implementation to the cloud. It also simplifies development, as developers only need to construct the processing pipeline once in the base class definitions and then focus on implementing functionalities for specific cases with their derived classes, avoiding repetitive pipeline construction and implementation whenever a new document type or analyzer service is added.
    \item \textbf{Integration of Non-Functional Requirements.} Developers integrate appropriate non-functional requirements into the corresponding objects to meet application needs for performance, availability, and cost. With Oparaca, non-functionality requirement enforcement, as shown in previous experiments, is achieved without any additional refinement effort from the developers.
\end{itemize}





\vspace{2mm}
\noindent
\colorbox{blue!10}{
\parbox{0.96\linewidth}{
\underline{\textbf{Takeaway}:} \emph{Oparaca accelerates development by abstracting low-level infrastructure concerns and automating runtime configurations through a high-level interface.
}}}
\section{Related Works}
\label{sec:related-work}

\subsection{Compute-Data Encapsulation}
Combining data and compute abstraction is an active research direction to deploy stateful applications with FaaS productively. We can classify studies on this front based on how the function can access the data.

\vspace{1mm}
\noindent\textbf{Unified compute-data abstraction.}
Many serverless platforms are designed to combine one or more functions and state data into unified deployment units such as ``actor'' ~\cite{spenger2024survey} or proclets \cite{ruan2023nu}. Functions and state data are co-located when executed so that functions can access the state data in local memory. Azure Entity Functions~\cite{azure_enfunc} that is based on the concept of virtual actor, Orleans~\cite{bykov2011orleans}. Kalix~\cite{kalix} uses CRDT~\cite{shapiro2011conflict} to replicate the state among functions. Similar to OaaS, our prior works ~\cite{lertpongrujikorn2023object,lertpongrujikorn2024object, lertpongrujikorn2024tutorial} and Nubes~\cite{marek2023nubes} also rely on the object-oriented concepts to encapsulate the function and data into unified deployments.

\vspace{1mm}
\noindent\textbf{Datastore abstraction.} The serverless platform provides a datastore API to the function for storing the state.
Cloudburst~\cite{cloudburst} offers stateful functions using a shared distributed key-value database. FAASM \cite{faasm} allows the function access to the shared memory via WASM. Crucial~\cite{crucial} allows a function to access the shared data via the DSO layer (distributed hash table). Boki~\cite{jia2021boki} enables stateful functions by providing API access to the distributed logging system. Beldi~\cite{beldi}, on the other hand, provides the database and transaction API to the state. YuanRong~\cite{chen2024yuanrong} offers a unified interface for the function to access the external database. Shredder~\cite{zhang2019narrowing} and Apiary~\cite{kraft2022apiary} enable the function to be executed within storage/database service in a stored-procedure manner.  Kalix~\cite{kalix} and Apache Flink Stateful Function (StateFun) \cite{statefun} proactively package the state within the invocation request payload and expect the modified state to be returned as part of the response payload.

Existing works, despite their diversity, focus mainly on data and compute encapsulation to enhance programmability and productivity, often neglecting non-functional requirements like performance and availability. OaaS fills this gap by introducing the new non-functional requirements interface and enforcing them by leveraging enriched information from the encapsulation with state-of-the-art solutions, as presented below.

\subsection{Non-functional Requirements Enforcement}

There is a rich body of research has been carried out to improve serverless execution latency. Most of them address the well-known cold-start problem \cite{wang18peeking, shadrad20serverless, ebrahimi2024cold}, which applications cannot easily resolve on their own.
Noticeable approaches include mitigating cold-start penalty \cite{du2020catalyzer, jia2021nightcore, liu2023faaslight} and sandbox recycling \cite{fuerst2021faascache, shadrad20serverless}.
Other efforts in the area focus on strengthening performance isolation \cite{agache2020firecracker, manco2017my} and proper resource allocation \cite{bhasi2022cypress, bilal2023great} to keep invocation executing at the desired speed.
Commercial cloud providers let applications manually configure for throughput through pre-allocation \cite{aws-lambda-provisioned-concurrency}, but this can be costly if the actual FaaS resource demand does not meet load estimation.  Real-time Serverless \cite{nguyen2019real} resolves the problem by allowing applications to dynamically scale to actual use under a predefined guaranteed invocation rate.

Enforcing the non-functional requirements becomes more complicated as applications evolve and become bigger. Many dedicated studies are addressing different aspects of the problem. Sequoia \cite{tariq2020sequoia} proposes a new QoS function scheduling and allocation framework. Real-time Serverless \cite{nguyen2019real, nguyen2023storm} extends the FaaS model to enable performance engineering through configurable guaranteed invocation rates. Aquatope \cite{zhou2022aquatope} proposes a QoS-and-uncertainty-aware resource scheduler for end-to-end serverless workflows. Astrea \cite{jarachanthan2022astrea} proposes an autonomous system that configures and orchestrates serverless analytic jobs. Pheromone \cite{yu2023following} replaces the traditional invocation-based workflow orchestration with a data-centric approach to enable locality exploitation across workflow execution.

Despite their significant benefits, the mentioned approaches address only limited aspects of FaaS applications. Furthermore, most rely on best-effort methods due to limited abstraction integration and cloud-developer coordination in current FaaS implementations and programming models (as presented in Section \ref{sec:problem}). This limits their practicality, as real-world applications often have multiple objectives and constraints \cite{nguyen2020motivating, nguyen2019real}. In contrast, OaaS's non-functional requirements API enables the enforcement of multiple objectives only through declaration with minimal refinement effort, allowing for simple and reliable application deployment.


\section{Conclusion}
\label{sec:conclusion}

In this paper, we introduced the Object-as-a-Service (OaaS) paradigm that offers a new cloud service abstraction that borrows principles of object-oriented programming to encapsulate application logic, data, and non-functional requirements into a unified deployment package.
The approach not only greatly simplifies native-cloud application development, but also enables requirements-driven cloud-developer coordination that opens the gate for many performance optimization opportunities. Moreover, OaaS relieves developers from the complexity of application fine-tuning to meet the desired QoS and deployment constraints. We also developed a prototype OaaS platform called Oparaca and evaluated it across various real-world applications and scenarios. The evaluation shows that Oparaca can enforce various application QoS with comparable resource efficiency versus other cutting-edge approaches while significantly reducing the time and effort required for cloud-native application deployment and development.

In the future, we plan to enhance Oparaca to support additional non-functional requirements (e.g., those listed in Table~\ref{tab:non-functional-requirements}). We will also expand its object configuration to give developers more flexibility in choosing data storage, execution, and orchestration implementations. This work serves as a starting point for several promising research directions. For instance, can Oparaca be extended across multiple data centers to leverage its high-level abstractions and non-functional requirement enforcement for addressing challenges in distributed systems, such as resilience and heterogeneity?


%

\section*{ACKNOWLEDGEMENT }
We would like to thank anonymous reviewers for their constructive feedback; and Chameleon Cloud for providing resources.
This project is supported by National Science Foundation (NSF) through CNS CAREER Award\# 2419588.

\balance
\bibliographystyle{ACM-Reference-Format}
\bibliography{references, oaas-references}


\end{document}